\documentclass[longauth]{aa}

\usepackage{graphicx}

\usepackage{txfonts,textcomp}
\usepackage{natbib}

\usepackage{upgreek}
\usepackage{array}
\usepackage{multirow}
\usepackage{multicol}

\newcommand{\thCO}{$^{13}$CO}
\newcommand{\CeiO}{C$^{18}$O}
\newcommand{\HCOP}{HCO$^{+}$}
\newcommand{\DCOP}{DCO$^{+}$}
\newcommand{\CCH}{C$_2$H}
\newcommand{\HCN}{HCN}
\newcommand{\HNC}{HNC}
\newcommand{\CN}{CN}

\newcommand{\Av}{$A_v$}
\newcommand{\Go}{$G_0$}

\newcommand{\Gon}{$G_0/n$}
\newcommand{\fe}{$f_{\rm e}$}

\newcommand{\NtHP}{N$_2$H$^{+}$}

\newcommand{\CNNtHP}{$W$(\rm{CN})/$W$(\rm{N}$_2$\rm{H}$^+$)}
\newcommand{\CNNtHPcd}{$N$(\rm{CN})/$N$(\rm{N}$_2$\rm{H}$^+$)}
\newcommand{\thCOHCOp}{$W$(\rm{$^{13}$CO})/$W$(\rm{HCO}$^+$)}
\newcommand{\CeiOHCOp}{$W$(\rm{C$^{18}$O})/$W$(\rm{HCO}$^+$)}

\newcommand{\CCHHNC}{$W$(\rm{C$_2$H})/$W$(\rm{HNC})}
\newcommand{\CCHHNCcd}{$N$(\rm{C$_2$H})/$N$(\rm{HNC})}
\newcommand{\CCHHCN}{$W$(\rm{C$_2$H})/$W$(\rm{HCN})}
\newcommand{\CCHCN}{$W$(\rm{C$_2$H})/$W$(\rm{CN})}

\newcommand{\Jone}{\mbox{(1--0)}}

\usepackage{orcidlink}
\newcommand{\orcid}[1]{\orcidlink{#1}}

\makeatletter
\DeclareRobustCommand{\HII}{%
  \mbox{H\check@mathfonts\fontsize\sf@size\z@\selectfont II}%
}
\makeatother

\makeatletter
\DeclareRobustCommand{\CII}{%
  \mbox{C$^{+}$}%
}
\makeatother

\makeatletter
\DeclareRobustCommand{\CI}{%
  \mbox{C\check@mathfonts\fontsize\sf@size\z@\selectfont I}%
}
\makeatother

\makeatletter
\DeclareRobustCommand{\HI}{%
  \mbox{H\check@mathfonts\fontsize\sf@size\z@\selectfont I}%
}
\makeatother

\usepackage{hyperref}
\hypersetup{colorlinks,allcolors=magenta}

\begin{document}

\title{Tracers of the ionization fraction in dense and translucent molecular gas: II. Using mm observations to constrain ionization fraction across Orion~B}

\titlerunning{Ionization fraction in Orion~B}

 \authorrunning{Be\v{s}li\'c}

\author{
  Ivana Be\v{s}lić\inst{\ref{LERMA/PARIS}} \orcid{0000-0003-0583-7363}
  \and Maryvonne Gerin\inst{\ref{LERMA/PARIS}} 
  \and Viviana V. Guzm\'an \inst{\ref{CHILE}} \orcid{0000-0002-0786-7307}
  \and Emeric Bron\inst{\ref{LERMA/MEUDON}} 
  \and Evelyne Roueff\inst{\ref{LERMA/MEUDON}} \orcid{0000-0002-4949-8562} 
  \and Javier R. Goicoechea\inst{\ref{CSIC}} 
  \and Jérôme Pety\inst{\ref{IRAM},\ref{LERMA/PARIS}} 
  \and Franck Le Petit\inst{\ref{LERMA/MEUDON}} 
  \and Simon Coud\'e\inst{\ref{WORC},\ref{CfA}} \orcid{0000-0002-0859-0805}
  \and Lucas Einig\inst{\ref{IRAM},\ref{GIPSA-Lab}}
  \and Helena Mazurek\inst{\ref{LERMA/PARIS}} 
  \and Jan H. Orkisz\inst{\ref{IRAM}} 
  \and Pierre Palud\inst{\ref{CRISTAL},\ref{LERMA/MEUDON}}
  \and Miriam G. Santa-Maria\inst{\ref{UF}} \orcid{0000-0002-3941-0360}  
  \and Léontine Ségal\inst{\ref{IRAM},\ref{IM2NP}} 
  \and Antoine Zakardjian\inst{\ref{IRAP}}
  \and S\'ebastien Bardeau\inst{\ref{IRAM}} 
  \and Pierre Chainais\inst{\ref{CRISTAL}} 
  \and Karine Demyk\inst{\ref{IRAP}} 
  \and Victor de Souza Magalhaes\inst{\ref{NRAO}} 
  \and Pierre Gratier \inst{\ref{LAB}} 
  \and Annie Hughes\inst{\ref{IRAP}} 
  \and David Languignon\inst{\ref{LERMA/MEUDON}} 
  \and François Levrier\inst{\ref{LPENS}} 
  \and Jacques Le Bourlot\inst{\ref{LERMA/MEUDON}} 
  \and Dariusz C. Lis\inst{\ref{JPL}} 
  \and Harvey S. Liszt\inst{\ref{NRAO}} 
  \and Nicolas Peretto\inst{\ref{UC}} 
  \and Antoine Roueff\inst{\ref{IM2NP}} 
  \and Albrecht Sievers\inst{\ref{IRAM}} 
  \and Pierre-Antoine Thouvenin\inst{\ref{CRISTAL}}
}

\institute{
  LUX, Observatoire de Paris, PSL Research University, CNRS, Sorbonne Universit\'es, 75014 Paris, France. \label{LERMA/PARIS}
  \and Instituto de Astrof\'isica, Pontificia Universidad Cat\'olica de Chile, Av. Vicu~na Mackenna 4860, 7820436 Macul, Santiago, Chile \label{CHILE}
  \and LUX, Observatoire de Paris, PSL Research University, CNRS, Sorbonne Universit\'es, 92190 Meudon, France. 
  \label{LERMA/MEUDON} 
  \and Instituto de Física Fundamental (CSIC). Calle Serrano 121, 28006, Madrid, Spain. \label{CSIC} 
  \and IRAM, 300 rue de la Piscine, 38406 Saint Martin d'H\`eres,  France. \label{IRAM} 
  \and Department of Earth, Environment, and Physics, Worcester State University, Worcester, MA 01602, USA \label{WORC}
  \and Harvard-Smithsonian Center for Astrophysics, 60 Garden Street, Cambridge, MA, 02138, USA. \label{CfA}
  \and Univ. Grenoble Alpes, Inria, CNRS, Grenoble INP, GIPSA-Lab, Grenoble, 38000, France. \label{GIPSA-Lab} 
  \and Univ. Lille, CNRS, Centrale Lille, UMR 9189 - CRIStAL, 59651 Villeneuve d'Ascq, France. \label{CRISTAL} 
  \and Department of Astronomy, University of Florida, P.O. Box 112055, Gainesville, FL 32611. \label{UF}
  \and Université de Toulon, Aix Marseille Univ, CNRS, IM2NP, Toulon, France,  \email{antoine.roueff@univ-tln.fr}. \label{IM2NP} 
  \and Institut de Recherche en Astrophysique et Planétologie (IRAP), Université Paul Sabatier, Toulouse cedex 4, France. \label{IRAP} 
  \and National Radio Astronomy Observatory, 520 Edgemont Road, Charlottesville, VA, 22903, USA. \label{NRAO} 
  \and Laboratoire d'Astrophysique de Bordeaux, Univ. Bordeaux, CNRS,  B18N, Allee Geoffroy Saint-Hilaire,33615 Pessac, France. \label{LAB} 
  \and Laboratoire de Physique de l'Ecole normale supérieure, ENS, Université PSL, CNRS, Sorbonne Université, Université de Paris, Sorbonne Paris Cité, Paris, France. \label{LPENS}
  \and Jet Propulsion Laboratory, California Institute of Technology,  4800 Oak Grove Drive, Pasadena, CA 91109, USA. \label{JPL}
  \and School of Physics and Astronomy, Cardiff University, Queen's buildings, Cardiff CF24 3AA, UK. \label{UC} 
}

\abstract 
{The ionization fraction (\fe $=n_\mathrm{e}/n_\mathrm{H}$) represents a fundamental parameter of the gas in the interstellar medium. However, estimating \fe\, relies on a deep knowledge of the underlying chemistry of molecular gas and observations of atomic recombination lines and electron-sensitive molecular emission, such as deuterated isotopologs of \HCOP\, and \NtHP, which are only detectable in the dense cores. 
Until now, it has been challenging to constrain the ionization fraction in the interstellar gas over a large areas because of limitations of observations of these tracers and chemistry models. }
{Recent models provided a set of molecular lines which ratios (intensities and column densities) can be used to trace \fe\, in different environments of molecular clouds. Here, we use a set of various molecular lines typically detected in the 3-4~mm range to constrain the ionization fraction across the Orion~B giant molecular cloud. In this work, we derive the ionization fraction for dense and translucent gas, and we investigate its variation with the density of the gas $n$ and the strength of the far-ultraviolet (FUV) radiation field \Go, with their ratio \Gon. }
{We present results for the ionization fraction across one square degree in Orion~B derived using analytical models and observational intensity and column density ratios of \CN\Jone/\NtHP\Jone, \thCO\Jone/\HCOP\Jone, and \CeiO\Jone/\HCOP\Jone\, in the dense and shielded medium (\Av~$\geq10$\,mag), and ratios of \CCH\Jone/\HNC\Jone, \CCH\Jone/\HCN\Jone, and \CCH\Jone/\CN\Jone\, in translucent gas ($2$\,mag\,$\leq$~\Av~$\leq6$\,mag).}
{ We find that the ionization fraction is in the range of $10^{-5.5}$, $10^{-4}$ for the translucent medium, and $10^{-8}$ to $10^{-6}$ for the dense medium. Our results show that the inferred \fe\, values are sensitive to the value of \Go, especially in the dense, highly UV-illuminated gas. We also find that the ionization fraction in dense and translucent gas decreases with increasing volume density ($f_e\propto n^{-0.227}$ for dense, and $f_e\propto n^{-0.3}$ in translucent gas), and increases with \Go, which is a consequence of how sensitive the emission of selected molecular lines (such as CN and \HCOP) is to the UV radiation field. In the case of the translucent medium, we do not find any significant difference in the ionization fraction computed from different line ratios. The range of \fe\, found in translucent gas implies that electron excitation of HCN and HNC becomes important in this regime. }
{In dense and shielded gas, we recommend using \CN\Jone/\NtHP\Jone\, to derive an upper limit on the ionization fraction \fe\, and \CeiO\Jone/\HCOP\Jone\, to set constraints on the lower limit. In a translucent medium, \CCH\Jone/\HNC\Jone\, serves as a good tracer of \fe. The moderately high \fe\, values found in translucent gas are consistent with the C$^+$/\CI/CO transition regime, while the values we find in the dense gas are sufficient to couple the gas with the magnetic field. }


\date{Received 09 01 2025; accepted 14 07 2025}

 \maketitle

\section{Introduction} 
\label{sec:intro}

The neutral interstellar gas is partially ionized \citep{McKee+1989}, which can be expressed through the ionization fraction, \fe, defined as: 

\begin{equation}
    f_e = \dfrac{n_\mathrm{e}}{n_\mathrm{H}},
\end{equation}

\noindent where $n_\mathrm{e}$ is the electron density, while $n_\mathrm{H} = n(\mathrm{H}) + 2n(\mathrm{H_2})$ is the density of hydrogen nuclei. The main mechanisms for ionization in the neutral interstellar gas are UV radiation from recently born stars and cosmic rays. Such presence of electrons and ions in the gas is crucial for initiating the molecular chemistry \citep{Herbst&Klemperer+1973, Oppenheimer&Dalgarno+1974}, particularly in the production of complex molecules \citep{Ceccarelli+2023}. In addition to their role in the chemistry of molecules in regions where they recombine with molecular ions such as \HCOP\, and \NtHP, electrons can be an important cause of molecular excitation in regions where $f_\mathrm{e} >10^{-5}$ \citep{Goldsmith+2001, SantaMaria+2023}. Recently, \citet{Gerin+2024} have shown that the excitation of the low-frequency transition of o-H$_2$CO at 4.8~GHz is sensitive to the ionization fraction under translucent gas conditions. The excitation of this line becomes overcooled (the excitation temperature of this transition becomes much smaller than the temperature of cosmic microwave background - CMB emission) only when \fe\, decreases below 10$^{-4}$. Moreover, through the processes of gas-magnetic field coupling \citep[for instance,][]{Basu&Mouschovias+1994, Padovani+2013, Pattle+2023}, the presence of charged particles such as electrons can also support molecular clouds against gravitational collapse.

The electron abundance depends on the local properties of molecular gas, particularly the gas density and extinction, parameters describing the ability of gas and dust to shield from UV radiation coming from young stars, cosmic ionization and the elemental abundance of species that provide electrons, such as atoms and PAHs \citep{Goicoechea+2009}. The main sources of electrons in the low-density, UV-illuminated gas are the photoionization of sulfur and carbon, whereas deeper in the cloud electrons are produced from ionization of H$_2$ from cosmic rays impact. Therefore, the ionization fraction varies across molecular cloud and ranges from $10^{-9}$ in the dense and deeply shielded medium, to $10^{-4}$ in the diffuse neutral gas at the surface of a cloud, where all carbon is expected to be ionized \citep{Williams+1998, Caselli+1998, Goicoechea+2009, Draine+2011, Salas+2021}. 

The exact value of \fe\, in the interstellar gas depends on the region, and it is generally challenging to measure. In models of molecular emission, \fe\, is often taken as a constant value, because \fe\, depends on parameters that are either not directly observable, or cannot be measured without deriving \fe\, first. The direct measure of the ionization fraction can be achieved using carbon and sulfur radio recombination lines (RRL) toward the dissociation fronts of bright photon-dominated regions (PDRs) \citep{Goicoechea+2009, Cuadrado+2019, Pabst+2024}. While this approach uses observations of free electron sources, it is limited toward a few regions where detection of RRL is possible \citep{Goicoechea+2021, Pabst+2024}. 

The alternative and most common approach is deriving \fe\, from abundances of molecules with chemistries involving recombination with electrons, such as deuterated species \citep[for instance, \DCOP\, and N$_2$D$^+$, among others:][]{Guelin+1977, Guelin+1982, Dalgarno&Lepp+1984, Caselli+1998, Caselli+2002, Fuente+2016}, through building stationary \citep{Williams+1998, Bergin+1999, Caselli+2002, Fuente+2016}, or time-dependent astrochemical \citep{Maret&Bergin+2007, Schingkedecker+2016, Goicoechea+2009}, and PDR models \citep{Goicoechea+2009}. The abundances of these molecules relative to their undeuterated forms, \HCOP\, and \NtHP, are used to constrain \fe\, through analytical formulae assuming a simplified chemical network and from computing the abundances of these species \citep{Caselli+1998, Caselli+2002, Miettinen+2011}. Recently, it has been shown that the ionization fraction and cosmic ray ionization rate are largely overestimated in cold and dense cores when information on the H$_2$D$^+$ column density is not used \citep{Redaelli+2024}. Using \DCOP/\HCOP\, and N$_2$D$^+$/\NtHP\, to compute \fe\, is limited to regions where these deuterated species are detected, i.e., cold cores in dense star-forming regions of the Milky Way.

All the methods mentioned above are observationally limited due to sensitivity challenges and are focused only toward specific regions within our Galaxy. Deriving the ionization fraction over an entire molecular cloud, from its illuminated and low-density regions to dense shielded and UV-illuminated parts, remains challenging and unexplored. With the current advancement in millimeter observing facilities, it is now possible to map several molecular features over entire clouds in relatively short integration times, opening new possibilities in understanding the ionization fraction in molecular gas. 

\citet{Bron+2021} (hereafter B21) took advantage of the available mm observations of the Orion~B giant molecular cloud \citep{Pety+2017} to derive several molecular line intensity and column density ratios sensitive to the ionization fraction from the analysis of extensive grids of 0-dimensional steady-state chemical models, without taking into account time dependence and geometrical effects. The results of this work are derived using a detailed chemical network from \citet{Roueff+2015} and the application of a random forest to predict which line ratio can serve as the best tracer of \fe.

As a result, B21 found a plethora of line ratios that can trace \fe\, more accurately than \DCOP/\HCOP, which is a consequence of the simplified chemistry often assumed in earlier works. B21 recommended using various intensity ratios to trace the ionization fraction, along with their corresponding empirical formulas for computing \fe, each representing statistical fits to the numerical models. However, the applicability of these models to real astronomical observations only started being exploited very recently \citep{Salas+2021}, and their comprehensive and more complete application remains to be fully explored, which is the scope of this work.

Orion~B is a giant molecular cloud located in the Orion constellation, east of the easternmost star of Orion's belt, Alnitak (top panel in Fig.\,\ref{fig:histograms}). At a distance of 410~pc \citep{Cao+2023}, this nearby cloud is known to host a few \HII\, regions, and their related PDRs, on its western side as a consequence of active star formation: the Horsehead nebula, an edge-on PDR created by the young, O-type star $\sigma$Ori; NGC~2024, a face-on PDR surrounding an \HII\, region illuminated by massive stars \citep{Bik+2003}; and NGC~2023, a reflection nebula.
Besides these regions of active star formation, the northern and eastern sides of Orion~B are quiescent places hosting prestellar cores, such as B~9, the Cloak in the East, and the northern Hummingbird filament. Thus, Orion~B represents a unique view of a giant molecular cloud where it is possible to investigate the ionization fraction in a span of different environmental conditions, from low-density regions, to dense cores and UV-illuminated gas.

In this work, we make use of several molecular lines in emission at 3~mm from the ORION-B large program \citep{Pety+2017}, combined with analytical models from B21, to derive the ionization fraction across the nearby Orion~B giant molecular cloud. Specifically, we apply models that consider Orion~B as a region consisting of dense and translucent gas. Here, we focus on the \mbox{\Jone} intensity ratios recommended in B21, which are: 

\indent $\bullet$ Dense medium: \CN/\NtHP, \thCO/\HCOP, and \CeiO/\HCOP. \\ 
\indent $\bullet$ Translucent medium: \CCH/\HNC, \CCH/\HCN, and \CCH/\CN. 

The paper is structured as follows: in Section\,\ref{sec:obs}, we describe the data set used in this work, including models to predict the ionization fraction from \cite{Bron+2021}. Next, we present our results for two gas regimes (dense and translucent) in Sections\,\ref{sec:dense} and \ref{sec:translucent}. Section\,\ref{sec:discussion} provides a discussion based on our findings. In Sec.\,\ref{sec:caveats}, we outline the caveats of our results and directions for future follow-ups, and finally, in Section\,\ref{sec:conclusions}, we summarize our work.

\section{Observations and analytical models} \label{sec:obs}

\begin{table*} 
\centering
\caption{The properties of the observed molecular lines used in this work.}
\label{tab:lines}
\setlength{\extrarowheight}{3.pt}
\begin{tabular}{l|cccc}
\hline 
Molecule  & $\nu_\mathrm{rest}$ [GHz] & $n_\mathrm{crit}$ [cm$^{-3}$] & $\theta_\mathrm{native}$ [''] & $\upsilon_\mathrm{int}$ [km/s] \\
\hline \hline 
C$_2$H\Jone    \,(h)  &   87.328    &  $1.0\cdot10^5$ &  28 & $[-40, -20]\cup[0, 20]$                \\
HCN\Jone       \,(h)  &   88.631    &  $2.7\cdot10^5$ &  28 & [-10, 20]                                \\
HCO$^+$\Jone          &   89.188    &  $4.5\cdot10^4$ &  28 & [0, 20]                                  \\
HNC\Jone              &   90.663    &  $7.0\cdot10^4$ &  28 & [0, 20]                                 \\
N$_2$H$^+$\Jone \,(h) &   93.173    & $4.1\cdot10^4$  &  28 & [-5, 25]                                 \\
C$^{18}$O\Jone        &   109.782   &  $4.8\cdot10^2$ &  23 &  [0, 20]                                \\
$^{13}$CO\Jone        &   110.201   &  $4.8\cdot10^2$ &  23 &  [0, 20]                                 \\
$^{12}$CN\Jone \,(h)  &   113.490   &  $2.4\cdot10^5$ &  23 &  $[-43, -30]\cup[-20, -7]\cup[0, 30]$  \\ \hline 
\end{tabular}
\tablefoot{The molecular lines used in this work are sorted by increasing rest frequency. We report the rest frequency, the critical density for collisions with H$_2$ at a kinetic temperature of 20\,K \citep{Shirley+2015, SantaMaria+2023, Zakardjian+2025}, native resolution, and the velocity range over which we integrate the main beam temperature to compute the integrated intensity. The letter h in brackets indicates that the corresponding molecular line has a hyperfine structure.}
\end{table*}

\subsection{IRAM 30-m observations towards Orion~B} \label{subsec:iram}

\begin{figure*}[t!]
    \centering
    \includegraphics[width=17cm]{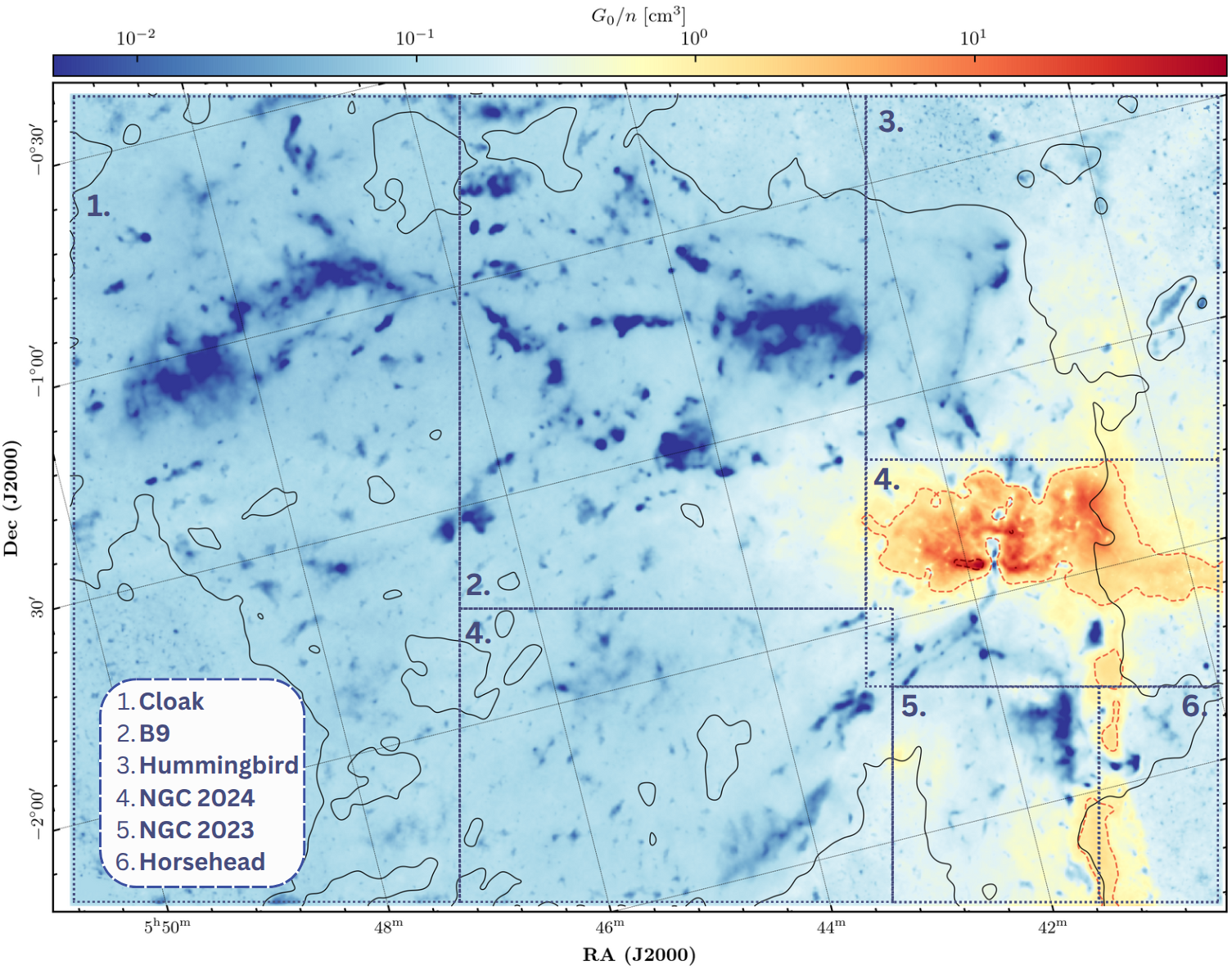}
    \includegraphics[width=17cm]{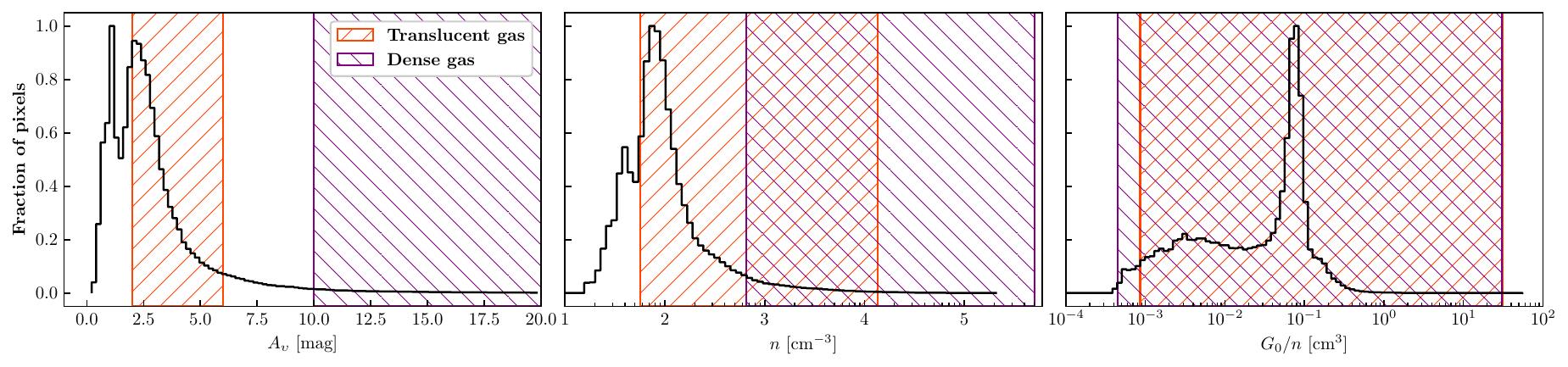}
    \caption{The ratio of the strength of FUV radiation field and volume density (\Gon) in Orion~B, and the distribution of pixels in dense and translucent gas. In the top panel, we show \Gon\, at 40\,arcseconds (0.08\,pc at the distance of the source) across Orion~B, where \Go\, is in Habing's units and $n$ in units of cm$^{-3}$. Black contour corresponds to \thCO\, integrated intensity of 0.5\,K\,km\,s$^{-1}$, indicating the boundaries of a GMC. Dashed contours correspond to values of \Gon\, of 1\,cm$^3$ (orange) and 20\,cm$^3$ (dark red). We divide Orion~B in six different regions. The bottom panels show histograms of normalized distribution of pixels having specific dust extinction (left panel), mean volume density (left panel) and \Gon\, (right panel). Orange and purple shaded regions correspond to values of pixels from translucent and dense medium, respectively.}
    \label{fig:histograms}
\end{figure*}

The Outstanding Radio-Imaging of OrionN B (ORION-B) Large Program carried out with the IRAM-30m telescope mapped a $\sim5$ square degrees area of the southern part of the Orion B molecular cloud (top panel in Fig.\,\ref{fig:histograms}). The Orion-B observations and data reduction are presented in detail in \cite{Pety+2017}. The data were obtained with the EMIR receiver and the FTS200 backend, resulting in a survey that covered the whole 3\,mm wavelength band (from 72 to 115.5~GHz, except for a small gap between 80 and 84 GHz) at high spectral resolution (195 kHz corresponding to 0.5\,km/s). The data reduction was done using the GILDAS software.

We list the properties of observed molecular lines (their rest frequency, the critical density, native resolution, and the velocity range used to compute the integrated intensity) used in this work in Tab.\,\ref{tab:lines}. We convolve the intensity maps to a common resolution of 40\,arcseconds, corresponding to $\sim0.08$\,pc at a distance of 410\,pc \citep{Cao+2023}.

\subsection{Intensities} \label{subsec:mom-maps}

We compute the integrated intensity ($W$ - moment 0) maps for the set of molecular lines used in this work, with their main properties shown in Tab.\,\ref{tab:lines}. First, we create a mask in the position-position-velocity cube to reduce the impact of the noise, following the method presented in \cite{Einig+2023}. We produce such a signal-to-noise mask for each molecule. The mask includes contiguous 3-dimensional regions with a signal-to-noise ratio (S/N) of the peak brightness of at least 2. After masking the data cube, we collapse the velocity axis along a specific velocity window (last column in Tab.\,\ref{tab:lines}) to ensure all the emission along the line of sight is captured. For computing $W$ of molecules that do not show a hyperfine structure, or for which the hyperfine structure is not resolved, as in the case of HNC, the integration window is fixed to the same velocity range as for \thCO. However, in the case of \HCN, \NtHP, \CCH\, and \CN\, that have a hyperfine structure (labeled with (h) in Tab.\,\ref{tab:lines}), resulting in the detection of individual components of the multiplet, we modify the velocity window to ensure that we have taken into account all the detected emission. We do not consider the presence of multiple velocity components \citep{Gaudel+2023, Beslic+2024}, or the existence of different gas layers along the line of sight \citep{Segal+2024}, that could be found in certain regions. Instead, we include the intensity of all the velocity components in our model.

Computing the integrated intensity of lines with a hyperfine structure requires an additional step, which is computing the intensity of the multiplet components used in analysis in B21. For instance, B21 did not consider the hyperfine splitting of \HCN\ and \NtHP, and we integrate the emission of all hyperfine components along the line of sight. For the remaining two molecules (\CCH\, and \CN), the hyperfine structure was taken into account. B21 used the intensity of the \CCH\ line at 87.328\,GHz, and the brightest component of the \CN\, multiplet at 113.490\,GHz. To properly measure the intensity of these specific components, we use the information on relative intensities of the components of the multiplets of \CCH\, and \CN\, (Tab.\,\ref{tab:hyperfine} in App.\,\ref{app:hyperfine}). We first compute the total intensities of \CCH\, and \CN\, and then multiply these by factors of 0.135 and 0.5, respectively. With this approach, we compute the intensities of \CCH(3/2, 1 $\rightarrow$ 1/2, 0) and \CN(5/2 $\rightarrow$ 3/2) used in B21, ensuring the minimization of the impact of the noise in the spectrum of these lines and possible cases when components or the multiplet are blended.  

Finally, to compute the column density ratios of \CN/\NtHP\, and \CCH/\HNC\, (hereafter \CNNtHPcd\, and \CCHHNCcd, respectively), we have assumed a local thermodynamic equilibrium (LTE) and an optically thin regime in which the column density is linearly proportional to the measured line intensity. When computing column densities, we assume the same excitation temperature $T_\mathrm{ex}=9.375\,$K for all species. These calculations are described in detail in App.\,\ref{app:cd}.

\subsection{Visual extinction, volume density and incident radiation field} \label{subsec:avg0}

Throughout this work, we use available maps of key parameters describing the interstellar gas across Orion~B, such as the maps of visual extinction \Av\, (Fig.\,\ref{fig:av}), the mean mass-weighted volume density ($n$) and incident radiation field \Go\,\citep[previously presented in][]{SantaMaria+2023}. The visual extinction is derived from dust maps obtained by Herschel and presented in \cite{Lombardi+2014}. We use the \Av\, map to define regions of translucent and dense gas (regions within the orange and purple contours in Fig.\,\ref{fig:av}) across Orion~B, assuming that \Av\, is a good tracer of the H$_2$ column density \citep{Planck+2014}.

The volume density maps of Orion~B are derived from column density maps and shown in \cite{Orkisz+2024}. These maps are a production of key properties from the probability density function (PDF) of volume density in each pixel, for instance, the mean volume density and the peak density. Here, we use the information on the mean volume density normalized by mass, assuming fully isotropic and unfragmented cloud. We compare \Av\, and $n$ across Orion~B to ensure that regions with high $n$ correspond to regions with high visual extinction, and we show this comparison in Fig.\,\ref{fig:n_vs_extinction} in App.\,\ref{app:maps}. 

The strength of the radiation field \Go, is computed based on the spectral energy distribution (SED) fits to the far-infrared (FIR)/submillimeter emission of dust grains heated by the interstellar radiation field \citep{SantaMaria+2023}. Throughout this study, we will investigate how the ionization fraction changes with the volume density, and mainly how it changes with the \Gon\,, which controls much of the physics and chemistry in the PDR gas \citep{Sternberg+2014}. We show a map of \Gon\, in the top panel of Fig.\,\ref{fig:histograms}.  

\subsection{Model predictions for the ionization fraction and selection of lines} \label{subsec:models}

\begin{table*}
\centering 
\caption{Parameters for the model predicted ionization fraction in dense and translucent gas from B21.} \label{tab:params-curves}
\label{tab:coeff}
\begingroup
\setlength{\extrarowheight}{3.5pt}
\begin{tabular}{l|ccccccc}
\hline
{\bf Dense medium }                         & \multicolumn{7}{c}{\bf Coefficients}                               \\
\hline
Tracer                                 & &  $a_0$            & $a_1$ & $a_2$ & $a_3$ & $a_4$ & $a_5$       \\
\hline \hline
$W(\rm CN\Jone)$/$W(\rm N_2H^+\Jone)$   & &   -6.192  & 0.5913  & -0.2320 & 2.217(-2) & 4.920(-2) & 5.057(-3)    \\
$W(\rm ^{13}CO\Jone)$/$W(\rm HCO^+\Jone)$ & & -7.891  &  1.249  & 2.2    & -3.222  &  1.556  & -0.2556            \\
$W(\rm C^{18}O\Jone)$/$W(\rm HCO^+\Jone)$ & & -7.081  &  1.597  & 0.3871 & -0.8281 & -0.1822 &  0.2346          \\
\hline 
{\bf Translucent medium }                   & \multicolumn{7}{c}{\bf Coefficients}                                 \\
\hline 
Tracer                                 & $f_\mathrm{max}$ & $a_0$ & $a_1$ & $a_2$ & $a_3$ & $a_4$ & $a_5$ \\
\hline \hline
$W(\rm C_2H\Jone)$/$W(\rm HNC\Jone)$   & -3.623  & -0.3627   & 0.8444   & -0.2051 & -3.929(-2) & 5.314(-2) &
  1.185(-2)       \\
$W(\rm C_2H\Jone)$/$W(\rm HCN\Jone)$   & -3.474  &   0.6598  &  0.6384  & -2.314  &  -1.681  &  -0.432  &  -3.817(-2)      \\
$W(\rm C_2H\Jone)$/$W(\rm CN\Jone)$    & 6.415   &  -10.77   &   0.3854 & 0.2471  &  0.8541  &   0.4505 &  6.662(-2)   \\ \hline  
\end{tabular}
\endgroup
\tablefoot{$W(\rm X)$ stands for the integrated intensity (moment-0) of the species X. Values that contain numbers in brackets should be multiplied by ten to the power of that number.}
\end{table*}

In the top panel of Fig.\,\ref{fig:histograms}, we show the boundaries of the Orion~B GMC with a black contour, defined using \thCO\Jone\, intensity detected across the region \citep[Figure\,1 in][]{Pety+2017}, corresponding to a $W(^{13}\mathrm{CO}(1-0))$ of 0.5\,K\,km\,s$^{-1}$. Within this region, \Av\, spans a range from 0.7\,mag to over 200\, mag. Following the models defined in B21, we compute the ionization fraction in two types of environments in Orion~B. These areas are defined based on the visual extinction: translucent ($2~\mathrm{mag} \leq A_v \leq 6~\mathrm{mag}$) and dense ($A_v \geq 10$\,mag) medium. By taking into account only pixels within the black contour in the top panel of Fig.\,\ref{fig:histograms}, we find that $16\%$ of them correspond to the diffuse medium ($A_v <2 $\,mag), $74\%$ of them are in the translucent medium ($2~\mathrm{mag}\leq A_v \leq6~\mathrm{mag}$), and $3\%$ of pixels are in the dense regions (\Av$\geq$10~mag). The remaining percentage ($7\%$) corresponds to a region between $6\,\mathrm{mag}< A_v <10$\,mag, which we do not analyze in this work, because this range of \Av\, contains gas that is still impacted by FUV radiation on the cloud, and therefore it does not represent the dense and shielded medium. In addition, we do not investigate the diffuse medium, since the ionization fraction in this region is expected to reach the values that one would derive from the relative carbon abundance (\fe$\approx10^{-4}$). In total, we compute the ionization fraction across a large portion ($77\%$) of Orion~B. 

The parametrization of the ionization fraction as a function of the line ratio ($x$) is given by a polynomial function (B21):
\begin{equation} \label{eq:xe}
  x(e) = 
\begin{cases}
  F(x), & \mathrm{Dense\,gas} \\
  f_{max} - \log( 1 + e^{-F(x)}), & \mathrm{Translucent\,gas}
\end{cases}
\end{equation}
where
\begin{equation}
  F(x) = a_0 + a_1 x + a_2 x^2 + a_3 x^3 + a_4 x^4 + a_5 x^5. \\
\end{equation}

The FUV radiation field heavily drives the chemistry in the translucent gas, whereas the dense gas becomes shielded against FUV radiation while becoming sensitive to cosmic rays. We note that the model of dense gas in B21 does not consider the stellar UV radiation field, and the analytical formula for dense gas is only valid for \Go=1. 

In the bottom panels of Figure\,\ref{fig:histograms}, we show the distribution of visual extinction (left panel), mass-weighted volume density (middle panel), and \Gon\, on the right panel. The distribution is shown for all pixels having $W(^{13}\mathrm{CO}(1-0))>0.5$\,K\,km\,s$^{-1}$. In these histograms, we show the ranges of \Av, $n$, and \Gon\, that are found in translucent and dense gas. The bottom left panel of Fig.\,\ref{fig:histograms} shows the criteria we use to define translucent and dense gas regions. In the case of $n$ and \Gon, the dense medium shows larger dynamical ranges of $n$ and \Go\, than the translucent gas. The translucent medium has a range of volume densities of ($60 - 1\cdot10^{4}$\,cm$^{-3}$) with a median value of 110\,cm$^{-3}$, whereas the range of volume densities in dense gas spans from $6.5\cdot10^2$ to $5\cdot10^5$\,cm$^{-3}$ with the median volume density of $3\cdot10^3$\,cm$^{-3}$. In the translucent gas, \Gon\, shows a range from $8\cdot10^{-4}$ to 31\,cm$^3$, and a range of $4\cdot10^{-4} - 30$\,cm$^3$ in the dense medium.

The coefficients defined in Equation\,\ref{eq:xe} for both mediums and the respective line intensity and column density ratios used in this work are given in Tabs.\,\ref{tab:coeff} and \ref{tab:params-curves}. In the work of B21, over 66 and 91 different line ratios (intensity and column densities for translucent and dense medium, respectively) were used to probe the ionization fraction. Here, we use the recommended line ratios in each medium to derive \fe. This selection was based on the sensitivity of molecular lines to electrons and the coefficient $R^2$ that measures the accuracy of the regression model to predict the ionization fraction (see, for instance, Table\,B6 in B21) and therefore the quality of the tracer (line intensity or column density ratio). In addition to this criterion, we also consider the detectability of molecular lines and focus on those typically used in galactic and extragalactic studies. In particular, we select to use the integrated intensity ratio and column density ratio of \CN/\NtHP\, for the dense medium. In addition to \CN/\NtHP, we also use \thCO/\HCOP{} and \CeiO/\HCOP{} to derive \fe\, in dense gas. For the translucent medium, we use the intensity and column density ratios of \CCH/\HNC, and will comment on the ionization fraction derived from the \CCH/\HCN{} and \CCH/\CN{} intensity ratios. 

\section{Results: Dense gas} \label{sec:dense}

\begin{figure}[t!]
    \resizebox{\hsize}{!}{\includegraphics{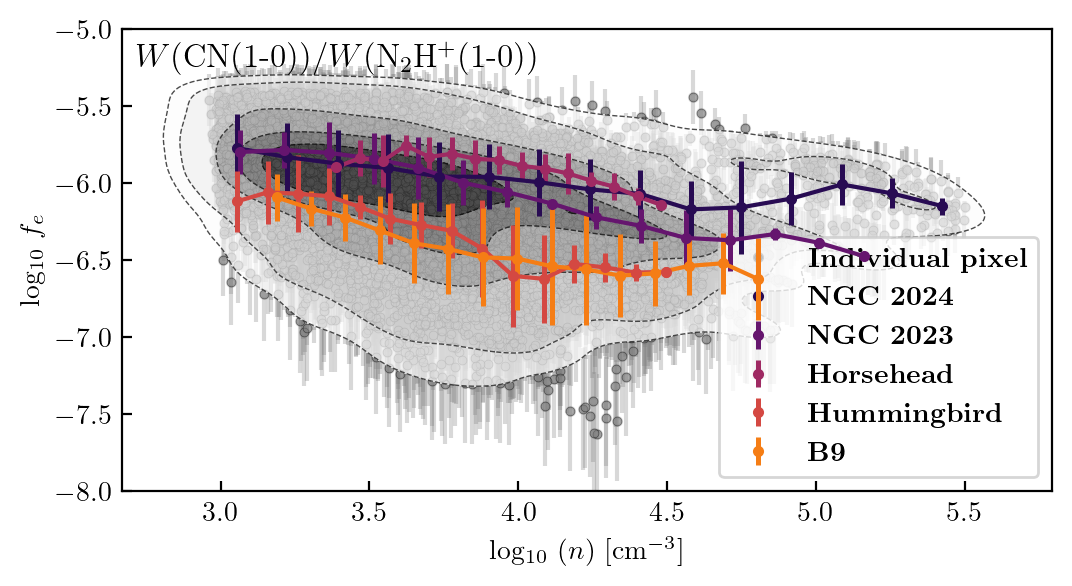}}
    \caption{Ionization fraction computed using the \CNNtHP\ ratio as a function of volume-weighted mean volume density, $n$. Contours correspond to the density of points of 1, 5, 25, 50, and 75 percent (from the most outer to the most inner contour). Colored dots are the binned trends of pixels from the different regions in Orion~B (top panel of Fig.\,\ref{fig:histograms}). The error bars show the weighted standard deviation of the points within each bin.}
    \label{fig:cn_n2hp_n}
\end{figure}

\begin{figure}[t!]
    \resizebox{\hsize}{!}{\includegraphics{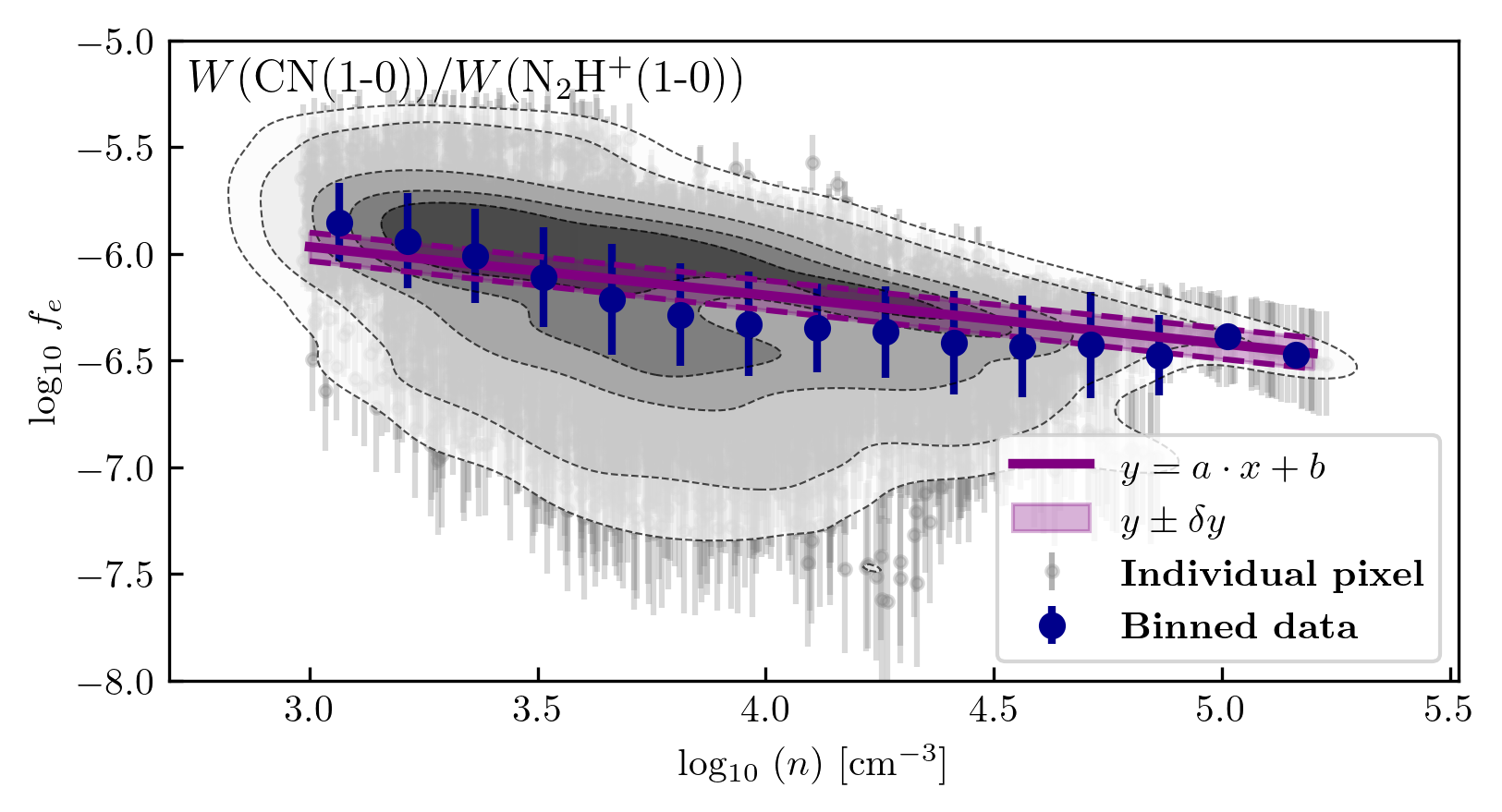}}
    \caption{The same as in Fig.\,\ref{fig:cn_n2hp_n}, but here we show binned trend of all data points except for NGC~2024 region. The purple line shows the linear fit and its uncertainty.}
    \label{fig:cn_n2hp_n_crb}
\end{figure}

All pixels satisfying the condition \Av$\geq10$\,mag are considered to belong to the dense medium shown within the purple contours in Fig.\,\ref{fig:av}. For clarity, we split the dense gas into a few regions shown in the top panel of Fig.\,\ref{fig:histograms}: NGC~2024, the Cloak, B~9, the Hummingbird filament, the Horsehead nebula, and the dense region between the Horsehead Nebula and NGC~2023. Each has different properties in terms of external radiation field and gas density. As seen from the top panel of Fig.\,\ref{fig:histograms}, these dense regions within Orion~B are differently exposed to stellar feedback, which we will also discuss here, given the fact that models of dense gas in B21 did not consider the impact of external radiation fields.

B21 recommended the \CNNtHP\, ratio as the best tracer of the ionization fraction in dense gas. On one hand, \CN\, is a typical tracer of PDRs, sensitive to UV radiation and often observed around young stars \citep{Pety+2017}. CN is also an important actor of nitrogen chemistry in dense and cold cores \citep{Hily-Blant+2013}. On the other hand, \NtHP\, is a typical tracer of dense and cold regions formed in the reaction between N$_2$ and H$_3^+$ and destroyed by recombination with electrons and reactions with CO. 

\subsection{Ionization fraction from the \texorpdfstring{CN(1-0)/N$_2$H$^+$}\,(1-0) line intensity ratio} \label{subsec:dense-intensities}

Firstly, we show the variation of \fe\, computed from \CNNtHP\, with the volume density in Fig.\,\ref{fig:cn_n2hp_n}. Each point represents one pixel from our map whose intensity ratio falls within the validity range defined by models in B21 (App.\,\ref{app:models}). In total, around 95$\%$ of \CN/\NtHP\, pixels (the fraction of the dense medium pixels where CN and \NtHP\, are detected) fall in the corresponding validity range (Fig.\,\ref{fig:dense_int_models}). The gray contours show the density of points. In addition, we bin the observed trend by region and volume density. These binned trends are shown in different colors for each region. The error bar of each binned point represents the measured uncertainties of points within each bin and also reflects the spread of the values.

The majority of data points have the ionization fraction around $10^{-6}$ (the dark grey region in Fig.\,\ref{fig:cn_n2hp_n}). We observe that the ionization fraction varies by region, and that \fe\, decreases with increasing volume density of the gas. The computed ionization fraction in dense gas spans from $10^{-5.5}$ in the lower-density regions ($n\approx10^3\,$cm$^{-3}$), and decreases toward the densest parts of Orion~B, where it reaches values between $10^{-7.5}$ and $10^{-6.5}$.  The highest \fe\, is present in NGC~2024, whereas the lowest \fe\, has been found in sightlines towards the Hummingbird and B~9.

Such a relationship between the ionization fraction and the volume density is expected from chemical models since the dense regions are more shielded against UV radiation, and recombination is more effective at high densities. Consequently, the emission of a UV-sensitive molecule, such as CN in this case, will become faint, while the emission of the density-sensitive molecule (\NtHP\, in this case) will increase, which will lower the \CNNtHP\ ratio and, therefore, also lower the observed ionization fraction (top panel in Fig.\,\ref{fig:dense_int_models}).

We further quantify the relation between the ionization fraction and volume density. Firstly, we bin all data points by volume density, excluding NGC~2024 region, because this region shows strong UV radiation, whose implications we will discuss later. These are shown in Fig.\,\ref{fig:cn_n2hp_n_crb}. Then, we fit the linear model ($log(f_e)=a\cdot log(n) +b$) to the binned values. Because we show and investigate the logarithmic values of ionization fraction and volume density in Fig.\,\ref{fig:cn_n2hp_n_crb}, the linear relation in log space corresponds to a power-law in linear space. We follow the prescription of such a linear regression from \cite{Segal+2024}. The fit is shown in purple in Fig.\,\ref{fig:cn_n2hp_n_crb}. The measured slope is $-0.227\pm0.003$ and the intercept is $-5.29\pm0.06$.

\begin{figure}[t!]
    \resizebox{\hsize}{!}{\includegraphics{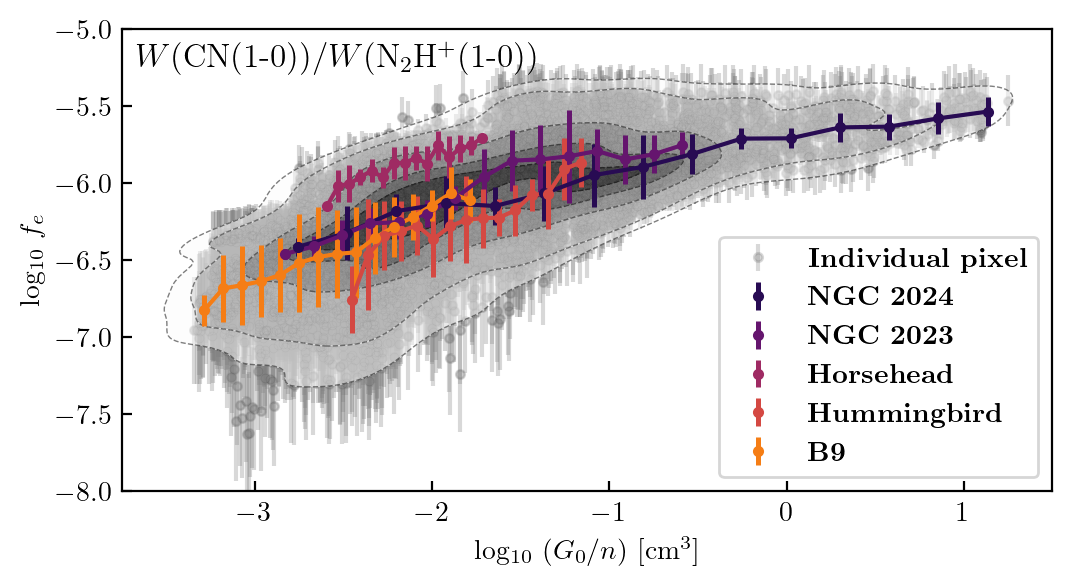}}
    \caption{The same as in Fig.\,\ref{fig:cn_n2hp_n}, but \fe\, is shown as a function of \Gon.}
    \label{fig:cn_n2hp_G0_G0n}
\end{figure}

Next, we show results on the ionization fraction computed using \CNNtHP\, as a function of the strength of the radiation field derived by the mass density-weighted mean volume density in Fig.\,\ref{fig:cn_n2hp_G0_G0n}. In this figure, \Gon\, (and \Go) increases to the right. Consequently, pixels showing moderate to high \Gon\, (yellow and red pixels in the top panel of Fig.\,\ref{fig:histograms}), also have the highest ionization fractions, whereas data points with low \Gon\, have moderate to low ionization fractions.

This result suggests that, besides the gas density, \Go\ has an impact on the computed ionization fraction of Orion~B, which is especially notable in regions of intense radiation field, in the West of Orion~B, such as NGC~2024. To investigate the impact of the local environment on \fe, we separately analyze pixels based on their location: B~9, the Cloak, the Hummingbird, NGC~2024, NGC~2023 and the Horsehead nebula. As seen in the top panel of Fig.\,\ref{fig:histograms} and Fig.\,\ref{fig:av}, these regions span a wide range in \Gon\, space, which makes it important to consider their properties when interpreting the \fe. To do so, similarly as in Fig.\ref{fig:cn_n2hp_n}, we bin data points within each region by \Gon, and include these binned trends in Fig.\,\ref{fig:cn_n2hp_G0_G0n}. The pixels in the Horsehead nebula show the lowest range of \Gon, whereas the broadest range in \Gon\, is found in NGC~2024 (almost four orders of magnitude). Here we also find that \fe\, depends on the region we analyze, and that regions with stronger \Go\, have higher \fe.

All regions shown in Fig.\,\ref{fig:cn_n2hp_G0_G0n} have the \fe\, between $10^{-7.5}$ and $10^{-5.5}$. The highest ionization fraction is found across NGC~2024 (up to \fe\, of $10^{-5.5}$), the region containing the densest gas and the strongest radiation field in our map. The ionization fraction of NGC~2023, the Horsehead, and the Hummingbird is similar (taking values from $10^{-6.5}$ and $10^{-5.75}$), whereas the lowest ionization fraction is found in B~9 ($\sim 10^{-7.5}$). Different \fe\, values in these regions can be explained by taking into account their characteristics. For instance, B~9 hosts cold, dense cores, and shows low \Gon, whereas NGC~2024 is the place of active star formation, with the presence of strong UV radiation. The Horsehead nebula is an edge-on PDR, and NGC~2023 is a reflection nebula surrounded by cold and dense gas. 

From Fig.\,\ref{fig:cn_n2hp_G0_G0n}, it is notable that the \fe\, depends on local properties, specifically the gas density and radiation field. This is especially observable in NGC~2024, where \Gon\, reaches the highest values of \Gon, and where we see the strongest variation of \fe. A strong radiation field can yield a stronger and deeper impact on the cloud, which can then enhance the emission of UV-sensitive molecules such as \CN\, and \HCOP\, \citep{Gratier+2017, Bron+2018}, and contributes to the increased number of ionizations. This interpretation can possibly explain the difference in binned trends between regions of low and high \Go, as shown in Fig.\,\ref{fig:cn_n2hp_G0_G0n}. However, as the gas becomes denser, the ionization fraction is expected to decrease since recombination reactions become more effective in the denser gas. We will further discuss our findings in the context of gas shielding and other impacts on observed line emission in the upcoming sections.

\begin{table*}[t!]
\centering
\caption{Tabulated median values of the ionization fraction across different regions in Orion~B: NGC~2024, NGC~2023, the Horsehead nebula, the Hummingbird filament, B~9 and the Cloak.}
\setlength\extrarowheight{6pt}
\label{tab:dense_gas_all_values}
{\begin{tabular}{l|c|cccc}
\hline
Region      & $(G_0/n)_\mathrm{median}$ & \multicolumn{3}{c}{$\log_{10} f_e$} \\ \hline
            & $10^{-3} [\mathrm{cm^3}]$               & $^{12}$CN\Jone/N$_2$H$^+$\Jone & $^{13}$CO\Jone/HCO$^+$\Jone & C$^{18}$O\Jone/HCO$^+$\Jone \\
\hline \hline 
NGC~2024    & 227                 & $-5.91_{-6.02}^{-5.81} $  & $-6.74_{-6.81}^{-6.57}$  & $-7.55_{-7.64}^{-7.40}$  \\ \hline
NGC~2023    & 44                  & $-5.96_{-6.14}^{-5.96}$   & $-6.58_{-6.75}^{-6.52}$  & $-7.35_{-7.47}^{-7.21}$  \\ \hline
Hummingbird & 43                  & $-5.86_{-5.94}^{-5.83}$   & $-6.41_{-6.58}^{-6.36}$  & $-7.19_{-7.38}^{-6.93}$  \\ \hline
Horsehead   & 16                  & $-5.99_{-6.12}^{-6.95}$   & $-6.91_{-6.93}^{-6.75}$  & $-7.42_{-7.45}^{-7.27}$   \\ \hline
B~9         & 3                    & $-6.46_{-6.57}^{-6.27}$   & $-6.53_{-6.64}^{-6.41}$ & $-7.14_{-7.27}^{-6.94}$  \\  \hline
Cloak       & 2                   &     /                      & $-6.49_{-6.61}^{-6.40}$ & $-7.11_{-7.22}^{-6.92}$  \\
\hline
\end{tabular} }
\tablefoot{We show the logarithm of the ionization fraction \fe\, derived using intensity ratios of \CNNtHP, \thCOHCOp, and \CeiOHCOp. We also tabulate percentiles (25th and 75th).}
\end{table*}

\subsection{Ionization fraction from \texorpdfstring{$^{13}$CO(1-0)/HCO$^+$(1-0)}\, and \texorpdfstring{C$^{18}$O(1-0)/HCO$^+$(1-0)}\, intensity ratios} \label{subsec:fe_dense_all}

We present results on  \fe\, in dense gas computed using \thCOHCOp\, and \CeiOHCOp. Optically thin CO isotopologs trace the cloud interior, particularly \CeiO\, \citep{Segal+2024}. \HCOP\, has a high critical density (Tab.\,\ref{tab:lines}) and is, similarly to CN, sensitive to UV radiation and present in PDRs \citep{Gratier+2017, Hernandez-Vera+2023}. \HCOP\, can trace the inner structure of clouds, as it can be formed in reactions initiated by cosmic rays \citep{Gaches+2019}. In addition, these lines have good S/N and are spatially extended across Orion~B \citep{Pety+2017}. 

\begin{figure*}
    \centering
    \includegraphics[width=17cm]{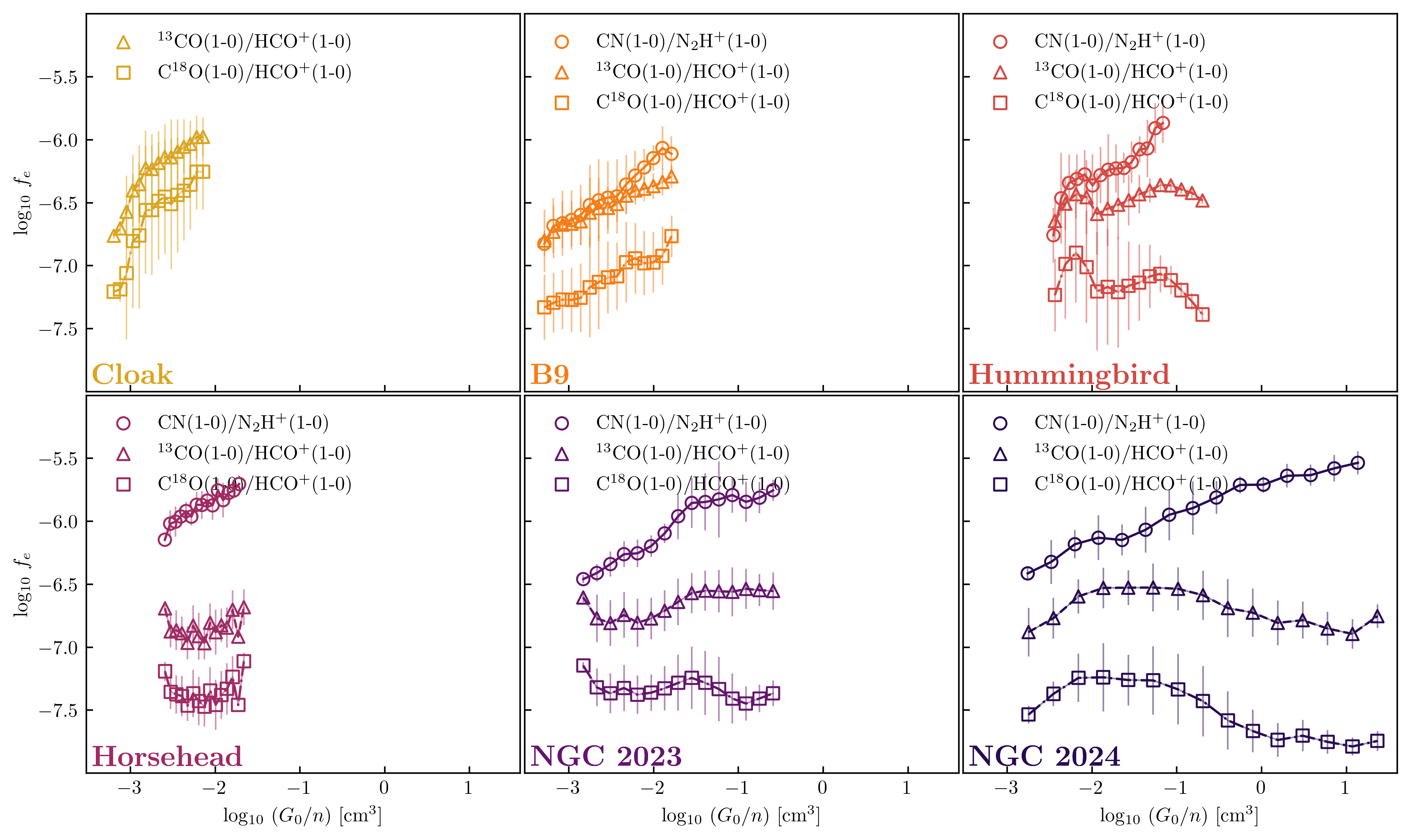}
    \caption{Ionization fraction in dense gas derived from intensity ratios of \CN\, and \NtHP, \thCO\, and \HCOP\, and \CeiO\, and \HCOP\, as a function of \Gon\, in six dense regions in Orion~B. We show binned trends by \Gon, and the errorbars correspond to the 25th and 75th percentiles.}
    \label{fig:dense_fe_all}
\end{figure*}

We show \fe\, derived from \thCOHCOp\, and \CeiOHCOp\, as a function of \Gon\, in Figs.\,\ref{fig:app-13co_hcop} and \ref{fig:app-c18o_hcop} in App.\,\ref{app:dense_others}. Similarly to the case of \CNNtHP, we have observed that \Gon\, impacts \fe\, derived from \thCOHCOp\, and \CeiOHCOp. To compare these findings, we show results on the ionization fraction computed using these three line intensity ratios for each region (reflecting different regimes of \Go\, as well, Sec.\,\ref{subsec:dense-intensities}) in Fig.\,\ref{fig:dense_fe_all}. We note here that \thCO, \CeiO, and \HCOP\, are also detected across the Cloak, which contains pixels with the lowest \Gon\, (around 2\,cm$^3$) across Orion~B.

We present median values of the ionization fraction derived from three line intensity ratios for each region considered as dense gas in Tab.\,\ref{tab:dense_gas_all_values}. The median value of the ionization fraction depends on the tracer and varies across regions. In particular, we note that the ionization fraction decreases with \Gon. We observe variations of \fe\, across each region computed from each line intensity and as a function of \Gon. The highest ionization fraction is derived from \CNNtHP\, in all regions except B~9, and the lowest \fe\, in dense gas in found when using \CeiOHCOp\, as a tracer. We also note the spread in values of \fe\, computed from these three line ratios. It seems that in regions with high \Go, such as NGC~2024, NGC~2023, the Horsehead, and even the Hummingbird, we observe a large spread in \fe, up to two orders of magnitude. 

In the bottom left panel of Fig.\,\ref{fig:dense_fe_all}, i.e., the case of NGC~2024, we see that \fe\, strongly depends on the selection of a tracer. For instance, the \fe\, derived from \CNNtHP\, is increasing with \Gon, whereas we observe the opposite behavior from the remaining two line ratios. A similar behavior is observed in the Hummingbird filament (top left panel in Fig.\,\ref{fig:dense_fe_all}). In the case of NGC~2023 and the Horsehead nebula (bottom middle and  right panels in Fig.\,\ref{fig:dense_fe_all}), \fe\, derived using all three line ratios behaves the same way with \Gon. The results on \fe\, are highly consistent for the case of B~9 and the Cloak (top middle and right panels in Fig.\,\ref{fig:dense_fe_all}) and do not depend on the choice of the line ratio. In these two regions, the ionization fraction increases with \Gon, ranging from $10^{-7.5}$ to $10^{-6}$. The slight difference between our results on \fe\, seen across B~9 and the Cloak is because the properties of these regions are closest to the model hypotheses presented in B21. In the case of B~9, the spread in \fe\, is larger than the uncertainties of these trends. We continue our discussion on the observed trends in Section\,\ref{sec:caveats}.

\section{Results: Translucent gas} \label{sec:translucent}

Translucent regions in Orion~B are shown within orange contours in Fig.\,\ref{fig:av}. To compute the ionization fraction of translucent gas, we have used integrated intensity ratios of molecules sensitive to FUV radiation, such as \CCH\, and CN and those whose emission scales with the column density: \HNC\, and \HCN.

\CCH, one of the simplest hydrocarbons, shows enhanced abundances in PDR gas \citep{Pety+2005, Beuther+2008, Cuadrado+2015, Goicoechea+2025}. The production of \CCH\, depends on the presence of C$^+$, one of the source of electrons in the ISM, which makes this molecule a good choice for probing the \fe\, in translucent gas. The production paths of HCN and HNC are coupled, and electrons are essential for their excitation in translucent region \citep{SantaMaria+2023}. At least 50$\%$ of total HCN and HNC emission across Orion~B is found in translucent gas \citep{SantaMaria+2023}. While HNC has an unresolved hyperfine structure, and its profile can be described using a single Gaussian function, the hyperfine structure of HCN\Jone\, is resolved in three separate hyperfine components. However, in most of the ORION-B field of view, the hyperfine components of HCN exhibit anomalous excitation \citep{Goicoechea+2022, SantaMaria+2023}, an effect not taken into account in B21 radiative transfer calculations, and which requires sophisticated calculations for deriving the HCN column density. Therefore, we chose to primarily focus on the analysis of \fe\, derived from the ratio of \CCH\, and \HNC\, (intensity and column density), and in the rest of this section, we will compare our findings with those computed from \CCHHCN\, and \CCHCN.

\subsection{The ionization fraction from the intensity ratio \texorpdfstring{\CCH(1-0)/\HNC(1-0)}\,} \label{subsec:trans-intensities}

\begin{figure}
    \resizebox{\hsize}{!}{\includegraphics{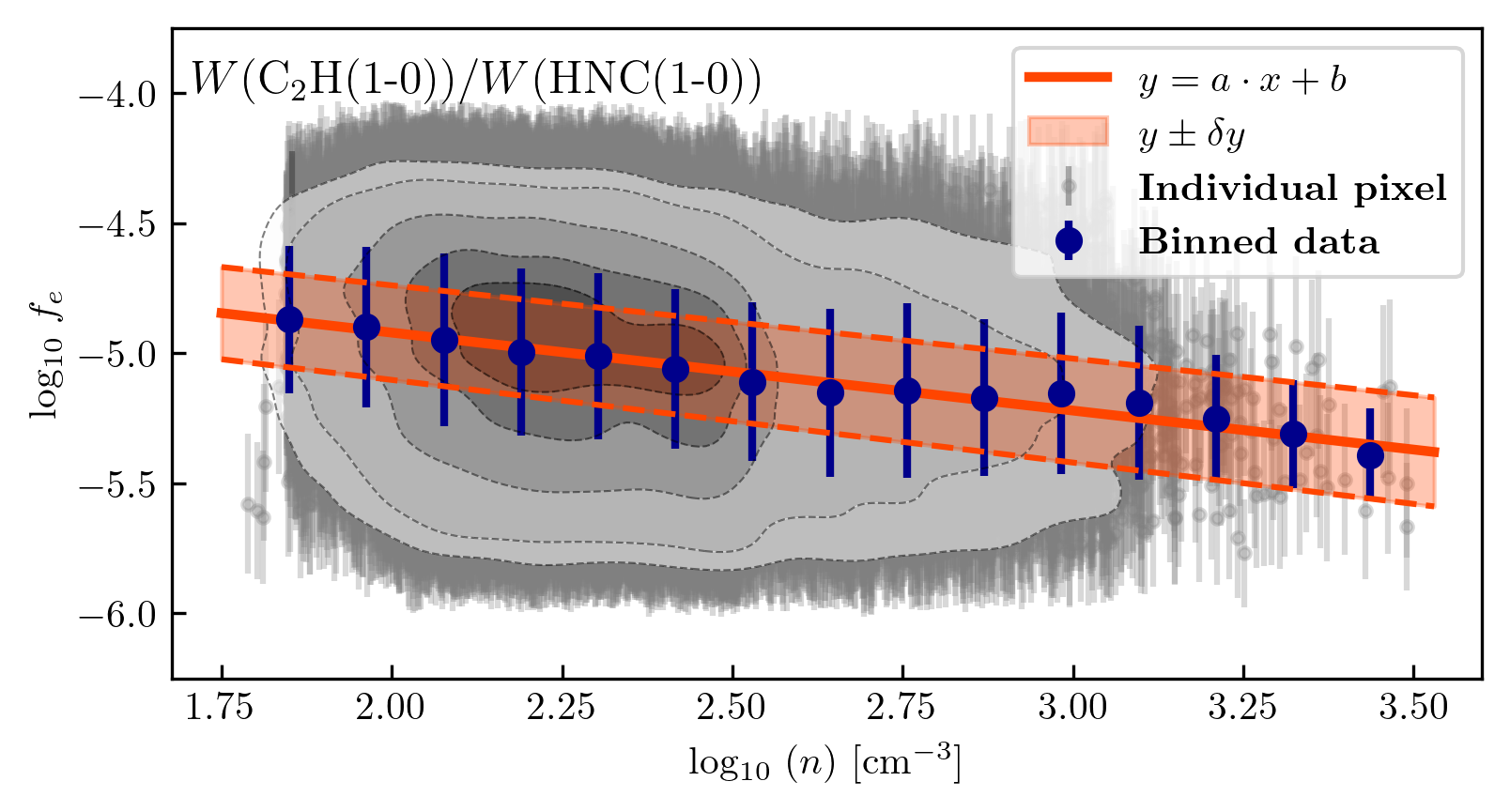}}
    \caption{Ionization fraction computed from \CCHHNC\, as a function of $n$ in translucent gas. Similarly as in Fig.\,\ref{fig:cn_n2hp_n}, we show the scatter of data points and the gray contours correspond to the density of data points. The binned trends are shown in dark blue. The error bars are computed in the same way as in Fig.\,\ref{fig:cn_n2hp_n}.}
    \label{fig:cch_hnc_n}
\end{figure}

\begin{figure}
    \resizebox{\hsize}{!}{\includegraphics{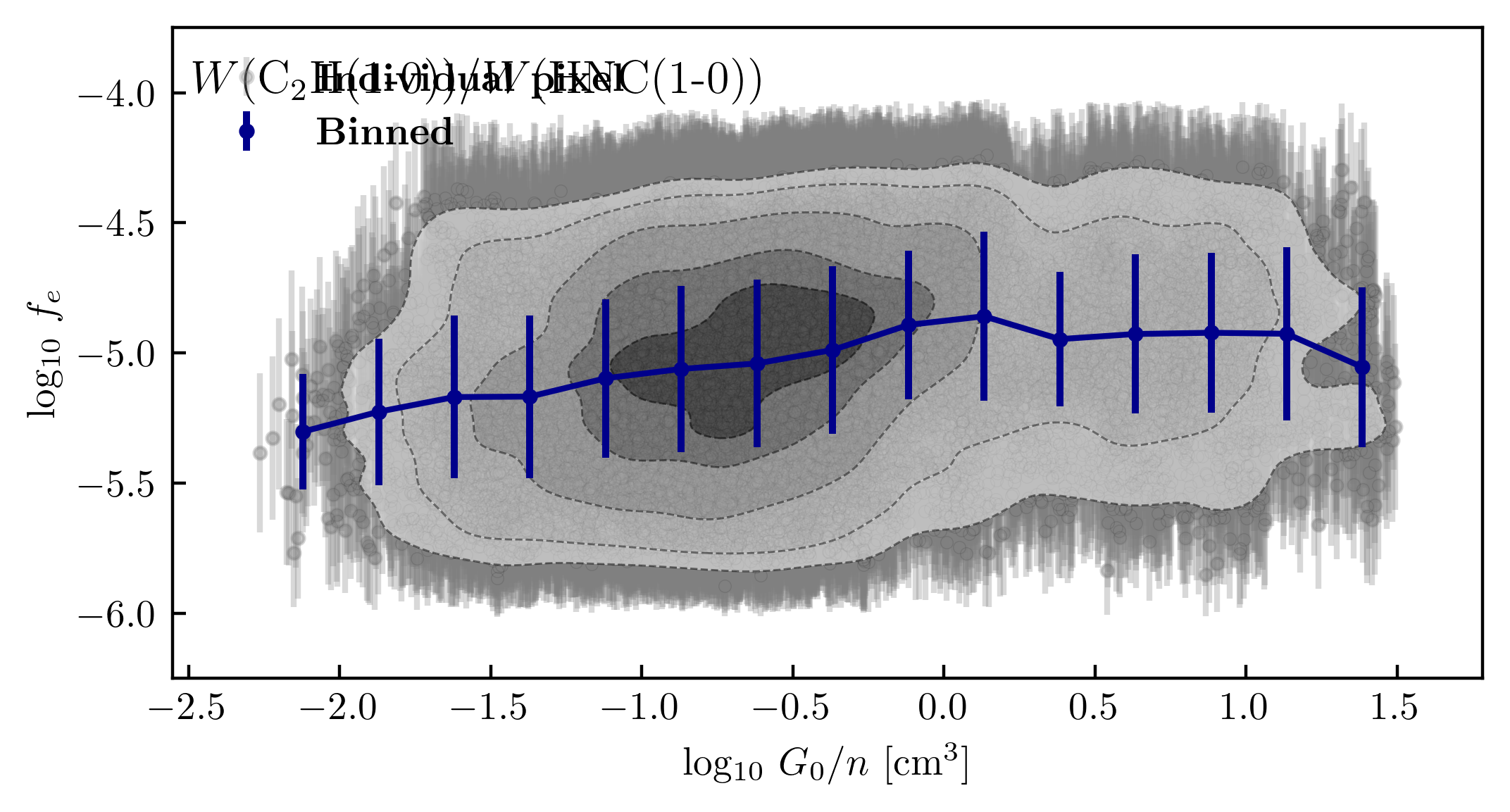} }
    \caption{The same as in Fig.\,\ref{fig:cch_hnc_n}, but for \Gon.}
    \label{fig:cch_hnc_Gon}
\end{figure}

We present results of \fe\, in the translucent medium computed using \CCHHNC\, as a function of $n$ and \Gon\, in Figs.\,\ref{fig:cch_hnc_n} and \ref{fig:cch_hnc_Gon} respectively. The density of data points is shown by grey contours, as in Fig.\,\ref{fig:cn_n2hp_n}. In both figures, the binned trend is shown in dark blue color. We find that the ionization fraction in translucent gas ranges between $10^{-5.5}$ and $10^{-4}$, which is higher than the ionization fraction found in dense gas (Sec.\,\ref{sec:dense}). The highest ionization fraction ($\approx 10^{-4.5}$) is found in the low-density regions ($n\approx10^2-10^{2.5}\,$cm$^{-3}$) in translucent gas (Fig.\,\ref{fig:cch_hnc_n}). The \fe\, decreases by 0.5\,dex toward higher volume densities, where it reaches $\approx10^{-5.25}$. These values of \fe\, found in translucent gas in Orion~B are between the typical values of \fe\, for the atomic ISM and for molecular regions where carbon is fully ionized ($10^{-4}$) and the typical \fe\, found in translucent gas \citep[$6\cdot10^{-5}$,][]{Snow+2006}.

In translucent conditions, it is expected that the ionization fraction decreases with increasing volume density. Such relation is described via power-law function, where: $f_e\propto\,n^{-1/2}$ \citep{McKee+1989, Caselli+1998}. Similarly as in dense gas, we fit a linear function to $\log_{10}$\fe--$\log_{10}n$, and show the result in orange in Fig.\,\ref{fig:cch_hnc_n}. We derive a slope of $-0.3\pm0.02$ and an intercept of $-4.32\pm0.14$.

The majority of pixels in Fig.\,\ref{fig:cch_hnc_Gon} have the lowest values of \Gon\, ($<1\,$cm$^3$) and are found in the entire map (Fig.\,\ref{fig:histograms}), while the pixels with \Gon\, greater than 1\,cm$^3$ are found in NGC~2024 and the Horsehead nebula. Here, we observe that the \fe\, increases from $\approx10^{-5.5}$ to $10^{-4.5}$ for \Gon\, going from $10^{-3}$ to $1\,$cm$^3$, after which \fe\, remains nearly constant. Overall, the higher values of the ionization fraction in translucent gas across Orion~B as compared with the dense gas regions suggest that carbon is partially ionized and sulfur could be fully ionized in this low density regime. The translucent regions then correspond to the transition zone from the ionized to the neutral carbon layer of a PDR. We further discuss this in Sec.\,\ref{sec:discussion_translucent}.

\subsection{The ionization fraction in translucent gas derived from \texorpdfstring{\CCH(1-0)/\HCN(1-0)}\, and \texorpdfstring{\CCH(1-0)/\CN(1-0)}\,} \label{subsec:fe_translucent_all}

We present results on \fe\, computed from \CCHHCN\, and \CCHCN\, in the translucent medium. The corresponding \fe\, as a function of \Gon\, is shown in Figs.\,\ref{fig:app-cch_hcn} and \ref{fig:app-cch_cn}. By using \CCHHCN\, and \CCHCN\, to trace \fe\, in translucent gas, we find that the value of the ionization fraction does not significantly vary depending on the tracer. The median \fe\, derived from these three line ratios is shown in Tab.\,\ref{tab:translucent_all}. We find that the median \fe\, ranges from $10^{-4.75}$ to $10^{-4.5}$. The lowest \fe, is derived using \CCHHNC, while the highest ionization fraction is calculated from \CCHCN. The highest scatter in \fe, is observed when using \CCHHCN, as a tracer (Fig.\,\ref{fig:app-cch_hcn}), where we also observe that several lines of sight reach values of $10^{-3.87}$, which is the saturation value defined by a coefficient $f_\mathrm{max}$ in Eq.\,\ref{eq:xe}, i.e., the fractional abundance of carbon relative to the total hydrogen content ($1.32\cdot10^{-4}$). 

\begin{table}[]
\centering
\caption{Tabulated median values of the ionization fraction in translucent gas using intensity ratios of \CCHHNC, \CCHHCN, and \CCHCN.}
\label{tab:translucent_all}
\setlength\extrarowheight{6pt}
\begin{tabular}{l|l}
\hline
Line intensity ratio $W_1/W_2$ & log$_{10}f_e$           \\
\hline \hline 
C$_2$H\Jone/HNC\Jone                     & $-5.04_{-5.23}^{-4.84}$ \\ \hline
C$_2$H\Jone/HCN\Jone                     & $-5.09_{-5.43}^{-4.86}$ \\ \hline
C$_2$H\Jone/CN\Jone                      & $-4.74_{-4.94}^{-4.54}$ \\
\hline
\end{tabular}
\tablefoot{We show median values of \fe\, and the 25th and 75th percentiles.}
\end{table}

\section{Discussion} 
\label{sec:discussion}

\subsection{Line intensity ratios versus column density ratios}
\label{sec:discussion_int_vs_cd}

Ionization fraction derived from the column density ratios for both mediums is presented in App.\,\ref{app:cd_results}. Here, we compare results on \fe\, derived from intensity and column density ratios (using the same molecular species) and justify our choice of using the line intensity ratio to compute the ionization fraction in Orion~B. In B21, models of \fe, derived from line intensity ratios are computed from chemical models and single-slab RADEX \citep{vanderTak+2007} models. These assumed uniform gas density, temperature, and electron abundance. Figs.\,\ref{fig:cn_n2hp_G0_G0n_cd} and \ref{fig:cch_hnc_cd} show how the ionization fraction computed from the column density ratios of \CN/\NtHP\, and \CCH/\HNC\, in dense and translucent gas, respectively, changes with the \Gon.

In dense gas, when using column density ratios to compute the \fe, we find that the ionization fraction also increases with the \Gon, as it is seen when using the intensity ratios of the same species (Fig.\,\ref{fig:cn_n2hp_G0_G0n}). The errors of data points are notably smaller than in the case of \fe\, derived from the intensity ratios, because of smaller model uncertainties (first two panels in Fig.\,\ref{fig:dense_int_models}). In this case, the \fe\, ranges from $10^{-7.5}$ in regions of low \Gon, and goes up to $10^{-5.5}$ in NGC~2024 and the Hummingbird. In translucent medium, the \fe\, derived from $N(\mathrm{C_2H})/N(\mathrm{HNC})$ does not vary with \Gon, and the median value is $\sim10^{-4.5}$.

Our results are consistent between the \fe\, derived from intensity and column density ratios. However, we chose to use the intensity ratios as tracers of the ionization fraction, because they can be directly derived from observations. In contrast, column densities are not directly observable, and their estimation from the observed line intensities requires assumptions on the physical conditions under which the lines are produced. In this case, we derive column densities from the observed line intensities assuming optically thin line emission and the LTE at a fixed excitation temperature for all considered species. 
The assumption of LTE stands for dense regime, whereas it needs to be revisited for the translucent regime. It is important to recognize that the observed molecular lines, despite their critical densities listed in Tab.\,\ref{tab:lines}, which are on the order of magnitude of $(10^{4}-10^{5})$ cm$^{-3}$, can still be produced below the critical density. Processes such as radiative trapping, where the line emission becomes optically thick, can reduce the effective critical density. Considering these effects, it may be more relevant to examine the effective critical density of these species, which accounts for such factors \citep{Shirley+2015}. For the lines analyzed in our study, the effective critical density is nearly two orders of magnitude lower than the conventional critical density (the density at which collisions dominate over radiative processes) reported in Tab.\,\ref{tab:lines}. Additionally, it is crucial to consider collisions with electrons, as they are the primary contributors to the excitation of molecules such as HCN and HNC at moderate densities \citep{SantaMaria+2023}.

\subsection{The ionization fraction in dense gas} \label{subsec:dense_discussion1}

The main ionization processes in dense molecular interstellar gas are related to cosmic rays, with additional more localized contributions from shocks created by protostars or supernovae. Cosmic rays ionize molecular hydrogen producing H$_2^+$, which in reaction with molecular hydrogen gives H$_3^+$, a molecular ion responsible for the production of several molecular ions, such as \HCOP\, and \NtHP\, \citep{Dalgarno+2006} through reactions with neutral molecular and atomic species, enabling an active ion-molecule gas-phase chemistry in dense shielded gas. In addition to the 
presence of low ionization potential species, such as sulfur and carbon, the presence of PAH is also crucial in understanding and determining the ionization fraction, since PAH-ions are formed straightforwardly through radiative attachment and may carry most of the electric charge.

In dense regions, we derive the ionization fraction using \CNNtHP, \thCOHCOp, and \CeiOHCOp. We find that typical values of the ionization fraction somewhat depend on the selected line ratio. General values of the ionization fraction in dense gas computed in this work range from $10^{-8}$ to $10^{-6}$, which are higher than the values ($5\cdot10^{-9}-10^{-7}$) reported in the literature for a cold dense core \citep[for instance,][]{Goicoechea+2009}. This difference comes from different regions probed and the different approaches to computing the ionization fraction. For instance, the value of $5\cdot10^{-9}$ reported in \cite{Goicoechea+2009} is found in a cold and dense, shielded core bright in \DCOP, whereas in our case, Orion~B contains several dense PDRs (dense regions impacted by strong UV radiation - NGC~2024, NGC~2023, the Horsehead, the Hummingbird). These regions contain dense gas, but they lie close to young stars that affect the surrounding medium. Moreover, B~9 and the Cloak are farther from the sources of UV radiation than the regions mentioned above, and we will discuss \fe\, which we found in these regions in the context of their properties.

We find a general dependence on the volume density, but the \Go\, in dense gas as well, as presented in Sec.\,\ref{sec:dense}. That is why we observe a positive correlation between \fe\, and \Gon. Our binning analysis revealed that different environments in Orion~B have different ionization fractions, which is expected because these regions show different properties. For instance, regions located in the west of Orion~B that are closest to the $\zeta$ Orionis star (Alnitak) have enhanced \Gon, driven by a strong radiation field. In addition to this, NGC~2024 shows the strongest star formation in the entire Orion~B cloud. The \CN\, emission, tracing the UV-illuminated gas \citep{Gratier+2017, Bron+2018} is bright in these regions.

It is important to note that the dense gas models presented in B21 do not consider variations of \Go, whereas the variation with the cosmic ray ionization is acknowledged. In that regard, dense regions with strong external UV radiation studied in this work do not fully correspond to the models from B21, and therefore these results on \fe\, should be treated with caution and as an upper limit. In the case of \CNNtHP, as shown in Fig.\,\ref{fig:dense_int_models}, our \CNNtHP\, measurements populate the central and right part of the validity range of the model of B21, missing a domain with a low \CNNtHP, line ratio that yields low \fe. In dense regions illuminated by strong FUV radiation, bright \CN, emission could be produced on the cloud surface, while \NtHP\, originates from inner parts of dense gas. Therefore, such regions would have a higher \CNNtHP, ratio than predicted by dense and shielded gas models, hence a higher value of the derived \fe, using B21 models.

\subsection{Ionization fraction as a local parameter in dense gas}
\label{subsec:dense_discussion2}

According to \cite{Hollenbach+1990}, the value of \Gon\, controls the structure of a PDR: in regions where \Gon\, is lower than $10^{-2}$\,cm$^{3}$, the self-shielding of H$_2$ is the dominant effect balancing photodissociation, while dust shielding becomes increasingly important for higher values of \Gon. Regions with \Gon\, above $0.1$\,cm$^3$ have a switch in the regime of the H/H$_2$ transition and in regions with highest \Gon\, (above 10\,cm$^3$), the radiation pressure supersonically drives dust through gas.

As previously discussed, \fe\, depends on the local properties of the gas. Therefore, we consider and compute the ionization fraction in several regions across Orion~B, which are considered as dense medium. There are six regions in total in Orion~B that host dense molecular gas (sorted from the easternmost of Alnitak): the Cloak, B~9, the Hummingbird, NGC~2024, NGC~2023, and the Horsehead nebula. Each of these regions has different properties in terms of the range of densities, \Go, structure and prestellar and protostellar cores, and can be divided into two categories: dense regions with low \Go\, (the Cloak, B~9, and the Hummingbird) and dense PDRs (NGC~2024, NGC~2023, and the Horsehead). In the case of the Cloak, we have not detected any significant CN and \NtHP\, emission in this region, and we will discuss results on \fe\, derived from other tracers used in this work in the next section.

\begin{itemize}
    \item B~9: This region has low \Go\, (see Tab.\,\ref{tab:dense_gas_all_values}), and shows the narrowest range in volume densities. Previous studies have shown that B~9 hosts prestellar and protostellar cores \citep[][]{Miettinen+2010, Miettinen+2012} that have dynamical evolution and it is an example of an isolated low-mass star-forming region. Nevertheless, the formation of cores in B~9 is probably influenced by nearby stars, as this region exhibits multiple velocity components observed in CO emission and its isotopologs \citep{Gaudel+2023}. We derive relatively low \fe\, values in B~9, of  $\sim 3\cdot10^{-7}$ with a scatter of a factor three.

    \item The Hummingbird: The Hummingbird filament is gravitationally stable \citep{Orkisz+2019} and, interestingly does not contain any YSOs, although some condensations and cores are detected as peaks of \CeiO\, emission. \fe\, shows a clear decreasing trend with increasing density in this filament. Overall \fe\, remains moderate but higher than in B~9, $\sim5\cdot10^{-7}$ with a scatter of a factor three.

    \item NGC~2023: NGC~2023 is a reflection nebula, and here we also include the dense region nearby, which explains variations in \fe\, across this area. In particular, we found a decreasing trend of \fe\, with increasing density and relatively low values of the ionization fraction in the coldest and densest ridge near to NGC~2023, similar to those found in B~9. The median \fe\, value is higher in NGC~2023 than in the previous regions, reaching about $7\cdot10^{-7}$.

    \item NGC~2024: NGC~2024 is probably one of the most interesting regions of the entire Orion~B cloud since here we find the most active star formation. In this case, we observe the largest variation in \Gon\, and in \fe. This is because NGC~2024 has a specific structure: it is composed of an \HII\, region, sourrounded by a dense PDR and the cold dusty filament (the Flame) located in front of the \HII\, region. Moreover, NGC~2024 hosts several protostellar cores, one of which drives a molecular outflow. The Flame filament is on the other hand cold and also a place of star formation. This structure of NGC~2024 can explain our results on \fe\, here: for instance, we have found lower \fe\, values along the Flame filament ($\sim4\cdot10^{-7}$), while higher values of the ionization fraction ($\sim2\cdot10^{-6}$) are located in regions of warm dust heated by stellar radiation.

    \item The Horsehead nebula: \cite{Goicoechea+2009} found a gradient of the ionization fraction in this region by investigating H$^{13}$CO$^+$ and \DCOP. The Horsehead nebula is an edge-on PDR, with a nearby cold dense core. The median value is intermediate between B~9 and NGC~2023 and close to that of the Hummingbird filament, about $5\cdot10^{-7}$. 
\end{itemize}

\subsection{Ionization fraction in translucent medium}
\label{sec:discussion_translucent}

In translucent medium, \fe\, takes values from $10^{-5.5}$ to $10^{-4.5}$ derived from \CCHHNC\, (Fig.\,\ref{fig:cch_hnc_n}), values between $10^{-5.75}$ and $10^{-3.87}$ when computed from \CCHHCN\, (Fig.\,\ref{fig:app-cch_hcn}), and $10^{-5.5}$ to $10^{-4.25}$ when traced by \CCHCN\, (Fig.\,\ref{fig:app-cch_cn}). In general, the results on \fe\, in translucent gas computed in this work are consistent among the different tracers. The median \fe\, derived in translucent gas from these three line intensity ratios, $\sim 2\cdot10^{-5}$, varies by less than a factor of two, and reaches a factor of four including the 25th and 75th percentiles (Tab.\,\ref{tab:translucent_all}). 

The range of values of the ionization fraction in translucent gas implies that the line emission analyzed in this work is produced in the \CII/\CI/CO layer, where ionized sulfur is an important contributor of electrons and carbon remains partially ionized \citep{Goicoechea+2021}. These conditions are consistent with the definition of translucent gas presented in \cite{Snow+2006}, in which some amount of carbon is ionized. 
That is why the expected ionization fraction in the translucent gas have values on the order of tenths of a percent of the fractional abundance of carbon, i.e., values between $10^{-5}$ and $10^{-4}$. \citet{Gerin+2024} have shown that the ionization fraction is very important for the excitation of the 4.8~GHz line of o-H$_2$CO. This line can have an excitation temperature lower than the (CMB) and be detected in absorption against the CMB for some specific conditions, and gets thermalized with the CMB for ionization fractions approaching 10$^{-4}$. Widespread absorption associated with extended \thCO\, emission is seen in translucent gas, which implies moderate values of the ionization fractions, as those independently derived in this work.

In addition, neutral carbon is detected throughout Orion~B \citep[bottom right panel of Fig.\,1 in][]{SantaMaria+2023, Ikeda+2002}, as well as emission from CO and \thCO\, \citep{Pety+2017}. Since the Einstein coefficient of the [\CI] fine structure line is low, the detected [\CI] emission originates from relatively large column densities of neutral carbon, comparable to those of CO. \cite{Ikeda+2002} found that atomic carbon represents between 10 and 20 per cent of all gas-phase carbon in Orion~B.

Furthermore, the high \fe\, values (\mbox{$\geq10^{-5}$}) found in low-density and high \Gon\, regions in Orion~B imply that the rotational level excitation of some molecules such as HCN and HNC \citep{Goldsmith+2017, Goicoechea+2022, SantaMaria+2023} is affected by collisions with electrons in the low-density gas. HCN and HNC 1--0 emissions are present in low-\Av\, gas; \cite{SantaMaria+2023} found that approximately $70\%$ of the total HCN emission comes from $A_v\leq8$\,mag and a correlation with the \CI\, emission. The photochemical models in \cite{SantaMaria+2023} yielded an ionization fraction of at least $10^{-5}$ in the translucent gas, sufficient for electron excitation to be efficient for HCN and HNC. 

\begin{figure*}
\centering
    \sidecaption
    \includegraphics[width=12cm]{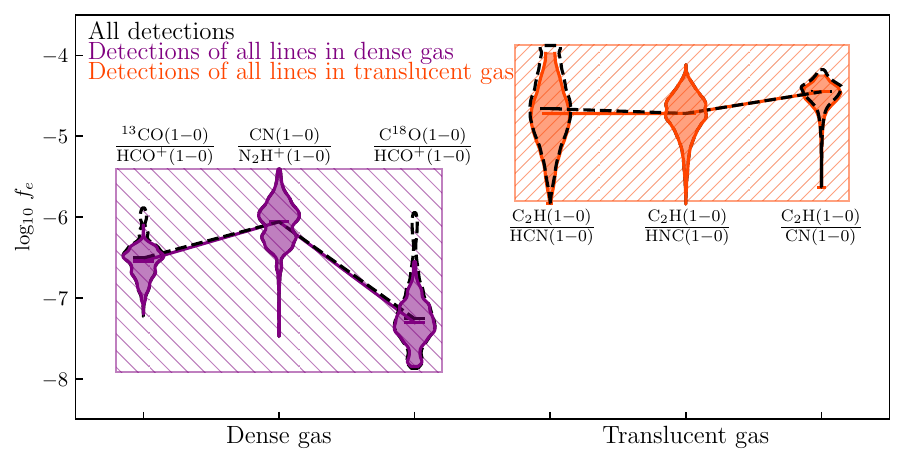}
    \caption{The distribution of the ionization fraction in dense medium computed from line ratios of \thCO/\HCOP, \CN/\NtHP\, and \CeiO/\HCOP\, (left part of the figure, purple shaded area). In the right part of this figure, we show results on \fe\, derived from \CCHHCN, \CCHHNC, \CCHCN\, in translucent medium (orange shaded area). Black dashed violins show distributions of all pixels that have detection in the corresponding intensity ratio. Purple violins show pixels only from regions where \CN/\NtHP\, is detected, whereas orange violins show the same for \CCHHNC. We also show the median for each of these categories as a horizontal line. }
    \label{fig:violin_all}
\end{figure*}

\section{Caveats}
\label{sec:caveats}

We further discuss our results in this section, with possible implications considering the models presented in B21 and results obtained in this work. Our results provide a certain range of the ionization fraction in dense and translucent gas. In both cases, we select tracers recommended in the work of B21. However, it is important to keep in mind the caveats of the models presented in B21 and their limitations. In addition, it is also important to further discuss the complexity of emission in Orion~B, and the molecular lines used in this work, including their sensitivity to environmental conditions such as gas density and stellar radiation.

\subsection{Model limitations and future improvements}

As previously stated, the models presented in B21 do not consider changes with \Go\, in dense gas. These models describe a dense gas region as a piece of gas shielded from the UV photons, where certain molecular species (such as \NtHP) are formed. However, Orion~B contains certain regions, especially those on its western side, that are both dense and highly affected by the UV field from young stars. Such specific conditions should be considered in future works, especially for the analysis of emission lines sensitive to FUV radiation.

\subsection{The reliability of tracers of \fe\, in Orion~B}
\label{subsec:dense_discussion3}

\subsubsection{Dense medium}

Here, we focus on discussing \fe\, derived from a set of different intensity ratios in dense gas. To interpret our results, we have to consider that each of these molecular lines are differently sensitive to gas conditions such as density and temperature. Our results (Fig.\,\ref{fig:dense_fe_all}) show that the highest ionization fraction is derived when using intensity ratio of \CNNtHP, while the smallest \fe\, is derived from \CeiOHCOp. In regions with low \Gon, such as B~9 and the Cloak, the ionization fraction shows the smallest variation among pixels and varies by less than 0.6\,dex between the different tracers, whereas in NGC~2024 for example, the variation in \fe\, depending on the choice of the tracer is up to 1.5\,dex. This result is supported by the fact that lines used to derive \fe\, in dense gas are not equally sensitive to \Go\, and the gas density which in turn will impact the estimated ionization fraction. For example, CN and \HCOP\, will contain a significant contribution from PDRs, while \CeiO\, and \NtHP\, only trace the dense and shielded gas. The line ratio \CNNtHP\, has a dense gas tracer in the denominator (the ratio of a line sensitive to UV/density-sensitive line), whereas in the case of \thCOHCOp\, and \CeiOHCOp\, the denominator is a PDR tracer (i.e., vice versa). This means that in regions where \CN\, and \HCOP\, get bright due to enhanced UV radiation, one line ratio will increase, while the other one will become lower.

Such a finding has a few implications. The first one is that we can trust our results in regions of low \Gon, especially in the Cloak, B~9, and the lowest \Gon\, regions of the Hummingbird, because the \fe\, derived from different tracers are overall consistent within less than 1\,dex. The difference arises as we move towards high \Gon\, regions because molecules sensitive to the intense UV field will have enhanced emission, increasing (lowering) the \CNNtHP\, (\CeiOHCOp) ratio. In such regions, CN and \HCOP\ become overbright and may come primarily from the surface of the cloud, whereas molecules such as \NtHP\ and \CeiO\ arise from the inner parts of the molecular cloud.

Another aspect that could help in the interpretation of our results lies in the following. Lines of sight in Orion~B are the superposition of multiple features (layers), associated with different velocity components in the spectrum of these molecular lines, most notably in CO isotopologs \citep{Gaudel+2023}, although \cite{Beslic+2024} found up to three velocity components in CN and \HCOP\, emission in NGC~2024. Furthermore, the observed spectrum of each molecular line is the result of radiative interaction of gas in multiple layers along the line of sight, such as dense cores and more diffuse medium. Such complex structures are not accounted by the simple chemical models used in B21. In this case, each molecule will differently trace such layers, and as a consequence, these layers will differently contribute to the total molecular emission \citep{Segal+2024}. By considering this scenario, it is probable that, for instance, in the case of \CeiOHCOp, these two emission lines probe and originate from somewhat different gas layers, as shown in the Horsehead nebula in \cite{Segal+2024}. Similarly, in the case of \CNNtHP, CN, although observed toward the entire NGC~2024 region could be mainly tracing  UV-heated gas, whereas \NtHP\, is coming in the cold and dense part.

Moreover, it is important to acknowledge the effect of opacity, in which case, the corresponding line emission can originate from the surface of a cloud, instead of probing the full line of sight. 
On the one hand, \thCO\, gets optically thick, and therefore its emission can trace only the outer parts of molecular clouds. Similarly, \CN\, and \HCOP\, emission can also get saturated \citep[as in the case of NGC~2024,][]{Beslic+2024}, but at higher densities than \thCO. On the other hand, \CeiO\, is known to be a good tracer of inner parts of molecular gas \citep{Pety+2017}, although it can be depleted in cold gas.

Considering all the discussion above, we provide a recommendation on the choice of the tracer of ionization fraction in dense gas. In all cases, we recommend the choice of molecular lines which approximately probe the same gas, where one of them is sensitive to the FUV radiation. In the case of regions with low \Gon\, (from $10^{-3}$\,cm$^3$ to $2\cdot10^{-2}$\,cm$^3$), we recommend using either of these three tracers, since \fe\, derived from these three line ratios nicely agree. In the case of regions with high \Gon\, (from $2\cdot10^{-2}$\,cm$^3$), we recommend to use \fe\, derived from \CNNtHP\, as an upper limit, and values of \fe\, derived from \CeiOHCOp\, as a lower limit.

In addition, it is important to bear in mind that \fe\, found in this work is derived from the total emission along the line of sight and provides a general overview of the ionization fraction across Orion~B. This means, as explained in Sec.\,\ref{sec:obs}, that we did not treat individual components of molecular emission coming from different velocity layers or distinguish between diffuse and dense gas along the line of sight (for example, filament versus envelope). At the moment, this type of analysis is beyond the scope of this work, but we plan to address this issue in future studies.

\subsubsection{Translucent medium}

Regarding the translucent gas, our results did not significantly vary based on the selection of a tracer, mainly since the excitation by electrons becomes important for HCN, HNC and CN. However, given the overlap with the validity range and resolved hyperfine structures of HCN and CN, including the anomalous excitation of HCN, we recommend the use of \CCHHNC\, for estimating the ionization fraction. 

\begin{figure*}[t!]
\centering
    \includegraphics[width=17cm]{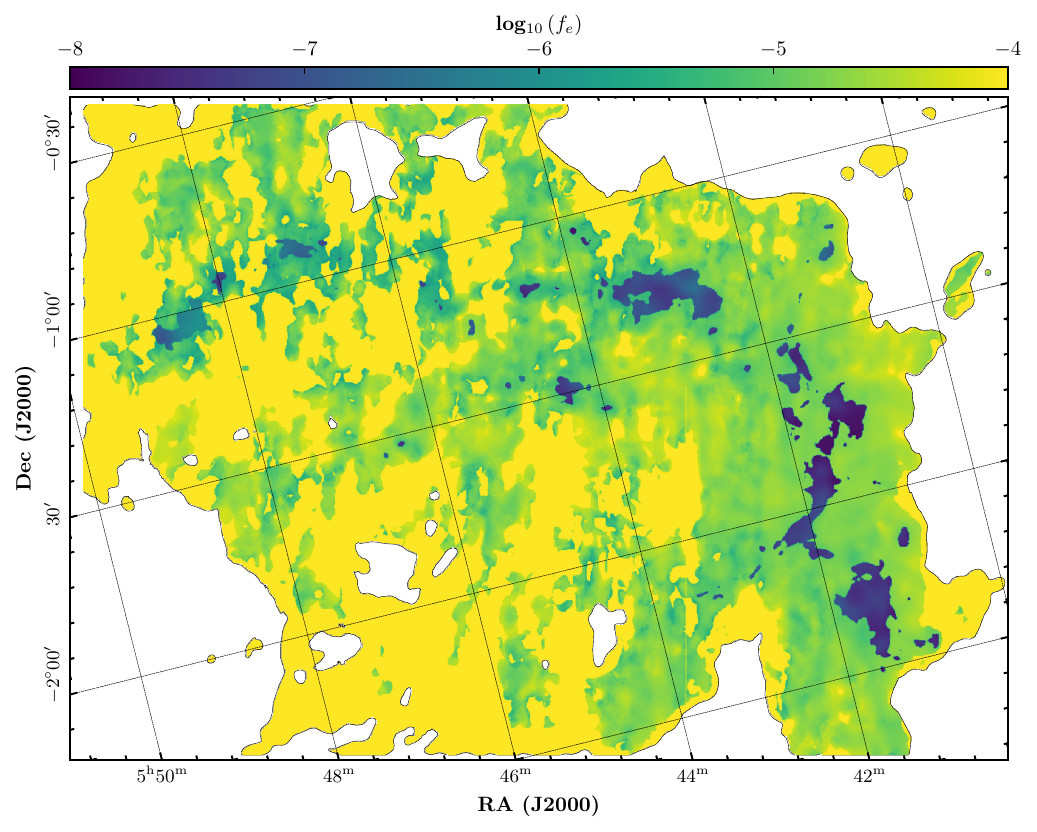}
    \caption{A map of ionization fraction across Orion~B. We show the \fe\, at 120\,arcseconds (0.24\,pc) derived using \CCHHNC\, in translucent gas and in regions of \Av\, between 6\,mag and 10\,mag, and \CeiOHCOp\, in dense gas. In diffuse medium (\Av$<2$) and in region where we do not detect any \CCHHNC, we set the ionization fraction to be $10^{-4}$. The black contours show the boundaries of Orion~B, defined from $W(^{13}\mathrm{CO}(1-0))$ emission.}
    \label{fig:fe_map}
\end{figure*}

\subsubsection{The impact of pixel selection criteria on the ionization fraction}

Finally, we comment on \fe\, computed from three different line ratios for both dense and translucent gas by considering the pixel selection effect. Firstly, we consider the dense medium. The overall \fe\, derived from \CNNtHP\, is different from that inferred from \thCOHCOp\, or from \CeiOHCOp\, for chemical reasons as discussed above, and also because the sensitivity threshold for detecting  these lines can also impact our results. Therefore, we compare the distribution of the ionization fraction derived from \CNNtHP, \thCOHCOp, and \CeiOHCOp\, in the left part of Fig.\,\ref{fig:violin_all}. In the same figure, we show the distribution of \fe\, from these line ratios obtained by taking into account exactly the same pixels from the map. \fe\, from \CNNtHP\, contains the lowest number of pixels, indicating that regions where both lines are detected probe the densest parts of Orion~B in comparison to \thCOHCOp\, and \CeiOHCOp. When we observe \fe\, from the same pixels, we see that the median \fe\, from \thCOHCOp\, and \CeiOHCOp\, is lower than in the case of taking into account all data points. This finding implies that \CNNtHP\, probes the densest parts of Orion~B, whereas \thCOHCOp\, and \CeiOHCOp\, are more spatially distributed, containing lower density pixels, that have a higher ionization fraction. 

\section{Summary} 
\label{sec:conclusions}

In this study, we have measured the ionization fraction across almost the entire Orion~B giant molecular cloud using IRAM\,30-meter observations from the ORION~B large program and analytical models to predict the ionization fraction from line intensity or column density ratios presented in \cite{Bron+2021}. We show the map of ionization fraction in Fig.\,\ref{fig:fe_map}. In the context of Orion~B, we analyzed the ionization fraction in translucent ($2\leq$~mag~\Av$\leq6$~mag) and dense gas (\Av$\geq10$~mag), and investigated its variations with the strength of the incident FUV radiation characterized by \Go\, and with the volume density $n$. Following the models and recommendations from \cite{Bron+2021}, we use the following tracers (\Jone\, transitions) of the ionization fraction:

$\bullet$ In dense gas, we use the intensity and column density ratios of \CN\, and \NtHP, and the intensity ratios of \thCO\, and \HCOP\, and of \CeiO\, and \HCOP. \\
\indent $\bullet$ In translucent regions, we computed \fe\, using intensity and column density ratios of \CCH\, and \HNC, and intensity ratios of \CCH\, and \HCN\, and \CCH\, and \CN.

\noindent Our key findings in the dense medium are:

\begin{itemize}
    \item The ionization fraction in dense gas is in the range from $10^{-7.75}$ to $10^{-6.5}$. 
    \item We find that the ionization fraction slightly decreases with increasing volume density, with a power-law slope of -0.227.
    \item When using \CNNtHP\, to compute  \fe, we find that regions with strong FUV field show higher ionization fraction than regions with low \Go. We do not observe such dependence on \Go\, when using the two other intensity ratios. 
    \item Our results on ionization fraction in the dense medium are consistent when using different intensity ratios in regions of weak FUV radiation (B~9, the Cloak, the Hummingbird, parts of NGC~2023), whereas in NGC~2024 and the Horsehead nebula the range of values of ionization fraction varies by an order of magnitude. 
    \item These results can be understood considering the sensitivity of CN and \HCOP\, to the UV radiation. In dense regions irradiated by strong UV radiation, \CN\, and \HCOP\, emission may result from the surface layers of FUV-illuminated clouds, whereas the \NtHP\, and \CeiO\, emission comes from the inner part of the clouds. 
\end{itemize}

\noindent The key results in translucent gas are:

\begin{itemize}
    \item The ionization fraction in translucent medium is found between $10^{-5.5}$ and $10^{-4}$. 
    \item Similarly to dense medium, we not observe a variation of \fe\, with $n$, with a somewhat steeper power-law slope of -0.3.
    \item \fe\, derived from \CCHHNC, \CCHHCN, and \CCHCN\, are consistent within the uncertainties. 
    \item The values of the ionization fraction found in this medium suggest that the molecular gas considered as translucent lies in the transition layer from ionized carbon to atomic carbon to CO, and where excitation with electrons becomes an important aspect of the excitation of HCN, HNC and CN rotational line emission.  
\end{itemize}

Considering all the above, we find the ionization fraction in the range of ($10^{-7.75}$, $10^{-6.5}$) in dense gas and in the range of ($10^{-5.5}$, $10^{-4}$) in translucent medium across Orion~B. We suggest using results derived from \CNNtHP\, as an upper limit of the ionization fraction, and \CeiOHCOp\, as a lower limit to the ionization fraction in dense gas. In terms of translucent gas, we recommend using the intensity ratio that mostly overlaps with the model validity range, which in our case is \CCHHNC. 

\begin{acknowledgements}

We thank the anonymous referee for useful comments that have improved the manuscript.
This work is based on observations carried out under project numbers 019-13, 022-14, 145-14, 122-15, 018-16, and finally the large program number 124-16 with the IRAM 30m telescope. IRAM is supported by INSU/CNRS (France), MPG (Germany) and IGN (Spain).
This work received support from the French Agence Nationale de la Recherche through the DAOISM grant ANR-21-CE31-0010, and from the Programme National ``Physique et Chimie du Milieu Interstellaire'' (PCMI) of CNRS/INSU with INC/INP, co-funded by CEA and CNES.
M.G.S.M. and J.R.G. thank the Spanish MICINN for funding support under grant PID2023-146667NB-I00. M.G.S.M acknowledges support from the NSF under grant CAREER 2142300.
Part of the research was carried out at the Jet Propulsion Laboratory, California Institute of Technology, under a contract with the National Aeronautics and Space Administration (80NM0018D0004).
D.C.L. acknowledges financial support from the National Aeronautics and Space Administration (NASA) Astrophysics Data Analysis Program (ADAP).

\end{acknowledgements}

\bibliographystyle{aa}
\bibliography{aa53706-25.bib}

\appendix

\section{Hyperfine structure of CN(1-0) and C$_2$H(1-0)} \label{app:hyperfine}

In Tab.\,\ref{tab:hyperfine}, we list information about the hyperfine structure of \CN\Jone\, and \CCH\Jone. The properties of CN\Jone\, are taken from \cite{Savage+2002}, while the information for \CCH\Jone\, is  from \cite{Padovani+2009}. The transition of CN\Jone\, is divided into two groups, each containing a multiplet of five hyperfine components. The brighter group is observed at $\sim113.5$\,GHz, while the second one can be observed at $113.12$\,GHz. Our observations specifically cover the brighter group of CN\Jone\, hyperfine multiplet, in which we detect four out of the five components (the first four rows related to \CN\Jone\, in Tab.\,\ref{tab:hyperfine}). On the other hand, \CCH\Jone\, consists of six hyperfine components. Our data set includes two of these components, 3/2, 2 $\rightarrow$ 1/2, 1, and 3/2, 1 $\rightarrow$ 1/2, 0.

\begin{table}[hbt]
\caption{Properties of the hyperfine splitting of \CN\Jone\, and \CCH\Jone.} \label{tab:hyperfine}
\begin{tabular}{l|ccc}
Molecule       & Transition, $F$           & $\nu$ [GHz] & $RI$ \\ 
\hline \hline 
               & 3/2 $\rightarrow$ 1/2       & 113.488142      & 0.1235             \\
               & 5/2 $\rightarrow$ 3/2       & 113.490985$^*$      & 0.3333             \\
$^{12}$CN(1-0) & 1/2 $\rightarrow$ 1/2       & 113.499643      & 0.0988             \\
               & 3/2 $\rightarrow$ 3/2       & 113.508934      & 0.0988             \\
               & 1/2 $\rightarrow$ 3/2       & 113.520414      & 0.0123             \\
\hline 
               & 3/2, 1 $\rightarrow$ 1/2, 1 & 87.284105       & 0.042              \\
               & 3/2, 2 $\rightarrow$ 1/2, 1 & 87.316898       & 0.416              \\
               & 3/2, 1 $\rightarrow$ 1/2, 0 & 87.328585$^*$       & 0.207              \\
C$_2$H(1-0)    & 1/2, 1 $\rightarrow$ 1/2, 1 & 87.401989       & 0.208              \\
               & 1/2, 0 $\rightarrow$ 1/2, 1 & 87.407165       & 0.084              \\
               & 1/2, 1 $\rightarrow$ 1/2, 0 & 87.444647      & 0.043              \\ \hline
\end{tabular}
\tablefoot{We show their transitions, rest frequencies, and relative intensities. Lines listed in Tab.\,\ref{tab:cd} are indicated with a star.}
\end{table}

\section{Column densities} \label{app:cd}

Assuming the LTE case and that our lines are optically thin (opacity, $\tau\ll1$), the column density of the upper state $\rm u$ of the considered molecular transition scales linearly with the integrated intensity: $N_\mathrm{u}=const\cdot\,W$. The full expression is as follows:

\begin{equation} \label{eq:nu}
    N_\mathrm{u} [\mathrm{cm^{-2}}] = 1.9436\cdot10^7\dfrac{\nu}{A_\mathrm{ul}} \cdot \left [ 1 - \dfrac{e^{(4.8\cdot\nu\cdot\,T_\mathrm{ex})} - 1}{e^{(4.8\cdot\nu\cdot\,T_\mathrm{CMB})} - 1} \right ]^{-1} \cdot W.
\end{equation}

\noindent In the above equation, $\nu$ is the frequency of the transition in units of 100\,GHz, $A_\mathrm{ul}$ is the Einstein coefficient of the transition in s$^{-1}$, $T_\mathrm{ex}$ is the excitation temperature in K, and $T_\mathrm{CMB}$ is the temperature of the cosmic microwave background emission of 2.73\,K. The total column density is then computed as:

\begin{equation} \label{eq:ntot}
    N_\mathrm{tot} = N_{\rm u}\dfrac{Q(T_\mathrm{ex})}{g_\mathrm{u}}\cdot\,e^{\dfrac{E_\mathrm{u}}{k_{\rm B}T_{\rm ex}}},
\end{equation}

\noindent where $Q(T_\mathrm{ex})$ is the partition function computed for excitation temperature $T_\mathrm{ex}$, $g_{\rm u}$ is the degeneracy of the upper level, $E_{\rm u}$ is the upper level energy, and $k_{\rm B}$ is the Boltzmann's constant. 

Tab.\,\ref{tab:cd} summarizes the parameters needed for computing the column densities of the corresponding species considered in this work. Einstein coefficients, the upper level energy, the degeneracy, and the partition function are taken from The Cologne Database for Molecular Spectroscopy \citep[\href{https://cdms.astro.uni-koeln.de/}{CDMS};][]{Muller+2001, Muller+2005, Endres+2016}. Assuming that the emission of these lines is subthermally excited \citep{Roueff+2021, Segal+2024}, we take the excitation temperature to 9.375\,K for this work.

Assuming the excitation temperature is the same for both species, in cases where we are not in the LTE regime, their column density ratio is almost independent of the choice of excitation temperature. We find that the column density ratio in this case changes by less than 10 percents from the one measured assuming the LTE case with the excitation temperature equal to the dust temperature. In the non-LTE case, assuming that the excitation temperature is different for each species for which we compute the column density ratio, we find that the column density ratio can change by up to 100 percents, but that in most of the cases the variation is less than 20 percents. Nevertheless, given the variation of \fe\, with respect to the column density ratios in Figs.\,\ref{fig:dense_int_models} and \ref{fig:translucent_int_models}, this difference does not introduce significant uncertainties in our measurements. We note, however, that appropriate measurements of column densities of the species analyzed in our work will be the topic of future studies.

\begin{table}[h!]
\caption{Parameters of selected molecules and $J=1-0$ transitions.} \label{tab:cd}
\setlength{\extrarowheight}{3.pt}
\begin{tabular}{l|llll}
\hline 
Molecule   & $A_\mathrm{ul}$ [s$^{-1}$] & $g_\mathrm{u}$ & $log(Q(T_\mathrm{ex}))$& $E_\mathrm{u}$ [K] \\
\hline \hline 
$^{12}$CN\Jone  & $0.119\cdot10^{-4}$  &  6  &  1.3579  &  5.447  \\
C$_2$H\Jone     & $0.154\cdot10^{-5}$  &  3  &  1.2852  &  4.197  \\
N$_2$H$^+$\Jone & $3.628\cdot10^{-5}$  &  27  &  1.6116  &  4.47   \\
HNC\Jone        & $2.690\cdot10^{-5}$  &  3  &  0.6683  &  4.35   \\ \hline       
\end{tabular}
\tablefoot{We provide information on Einstein coefficient, degeneracy, partition function computed assuming the excitation temperature of $T_\mathrm{ex}=9.375\,$K, and the energy of the upper level.}
\end{table}

\begin{table*}
\centering 
\caption{Parameters for the model predicted ionization fraction in dense and translucent gas from B21 derived using column density of molecular lines.}
\label{tab:coeff_cd}
\begingroup
\setlength{\extrarowheight}{3.5pt}
\begin{tabular}{l|ccccccc}
\hline
{\bf Dense medium }                         & \multicolumn{7}{c}{\bf Coefficients}                               \\
\hline
Tracer                                 & &  $a_0$            & $a_1$ & $a_2$ & $a_3$ & $a_4$ & $a_5$       \\
\hline \hline
$N(\rm CN\Jone)$/$N(\rm N_2H^+\Jone)$   & & -7.561    &  0.7191 & 4.245(-2) & -5.030(-2) & 1.433(-3) & 1.432(-3)      \\
\hline 
{\bf Translucent medium }                   & \multicolumn{7}{c}{\bf Coefficients}                                 \\
\hline 
Tracer                                 & $f_\mathrm{max}$ & $a_0$ & $a_1$ & $a_2$ & $a_3$ & $a_4$ & $a_5$ \\
\hline \hline
$N(\rm C_2H)$/$N(\rm HNC)$   & -3.525  &  -1.265   &  0.9225  & -0.0424 &  -0.1323 &  0.0176  &  1.387(-2)      \\
\hline  
\end{tabular}
\endgroup
\tablefoot{$N(\rm X)$ is the column density of the species X. Values that contain number in brackets should be multiplied by ten to the power of that number.}
\end{table*}

\section{Visual extinction as a tracer of volume density in Orion~B}
\label{app:maps}

\begin{figure*}[t!]
    \includegraphics[width=0.95\textwidth]{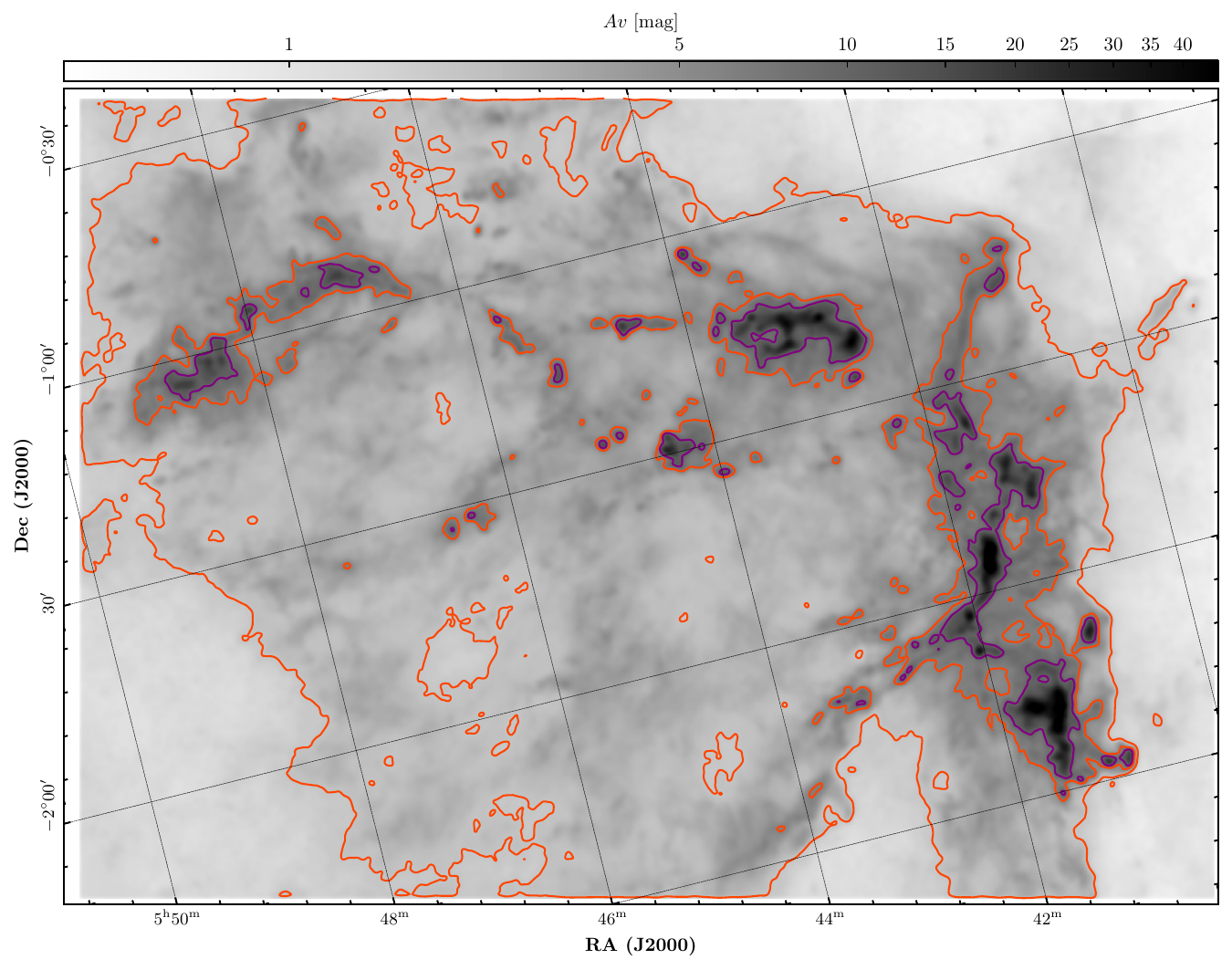}
    \caption{Dust visual extinction across Orion~B. This map was derived from dust column density \citep{Lombardi+2014}. Orange contours show regions of translucent gas, where \Av\, is in the range from 2\,mag to 6\,mag. Purple contours show regions of dense gas.} \label{fig:av}
\end{figure*}

\begin{figure}[hbt!]
    \resizebox{\hsize}{!}{\includegraphics{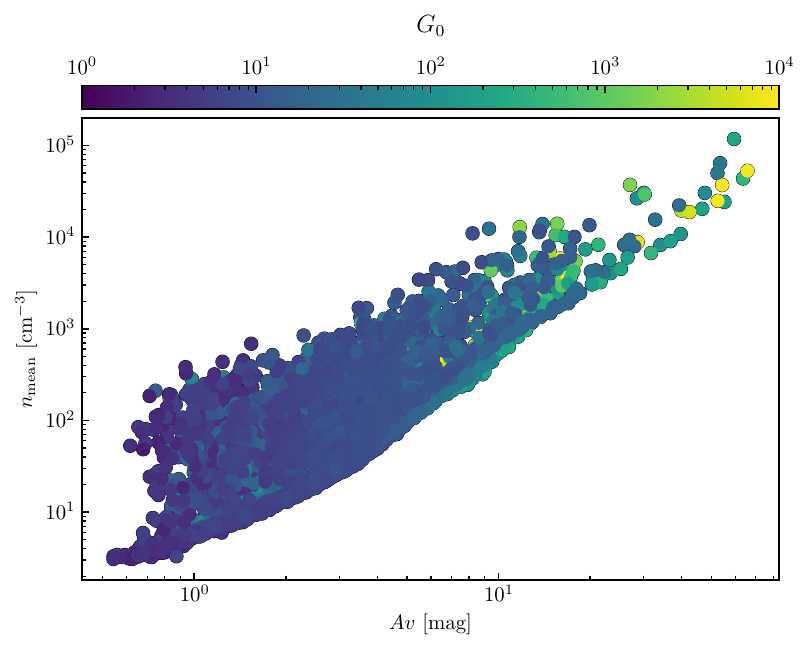}}
    \caption{Mean mass density-weighted volume density as a function of visual magnitude. We color-code each point by a corresponding strength of radiation field ($G_0$).} \label{fig:n_vs_extinction}
\end{figure}

We use the map of the dust visual extinction, \Av\, from \citet{Lombardi+2014}, shown in Fig.\,\ref{fig:av}, to define regions of dense and translucent molecular gas in Orion~B. In this figure, translucent and dense gas regions are found within the orange and purple contours, respectively. Since we are investigating how the ionization fraction varies as a function of \Go\, and volume density $n$, we will explore the relationship between \Av\, and $n$. For further information on the derivation of volume density maps of Orion~B, we direct the reader to \cite{Orkisz+2024}. In Fig.\,\ref{fig:n_vs_extinction} we present the mean mass density-weighted volume density \citep{Orkisz+2024} as a function of visual extinction with points color-coded by their \Go\, values. The observed scatter indicates a positive correlation between \Av\, and $n$.

\section{Models and validity ranges} \label{app:models}

\subsection{Dense medium}
\label{app:dense}

\begin{figure}
    \centering
    \includegraphics[scale=0.88]{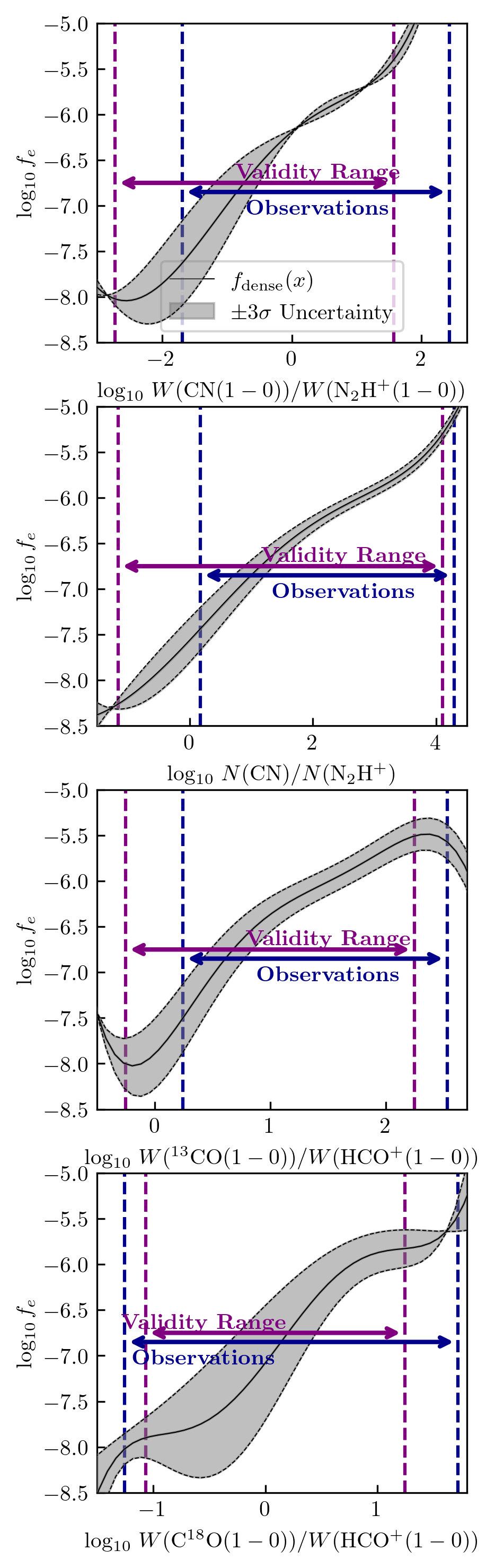}
    \caption{Ionization fraction as a function of selected line intensity ratios for dense gas. The range of observed line ratios is shown in blue and the validity range is in purple. We show the analytical function as a black solid line showing relation between \fe\, (y-axis) and intensity ratios (x-axis) of \CN/\NtHP{} (top panel), \CNNtHPcd\, (second panel), \thCO/\HCOP{} (third panel), and \CeiO/\HCOP{} (bottom panel). The analytical functions (Eq.\,\ref{eq:xe}), its $3-\sigma$ variance (grey shaded area) and the validity ranges are taken from \cite{Bron+2021}.}
    \label{fig:dense_int_models}
\end{figure}

\begin{figure}
    \centering
    \includegraphics[scale=0.88]{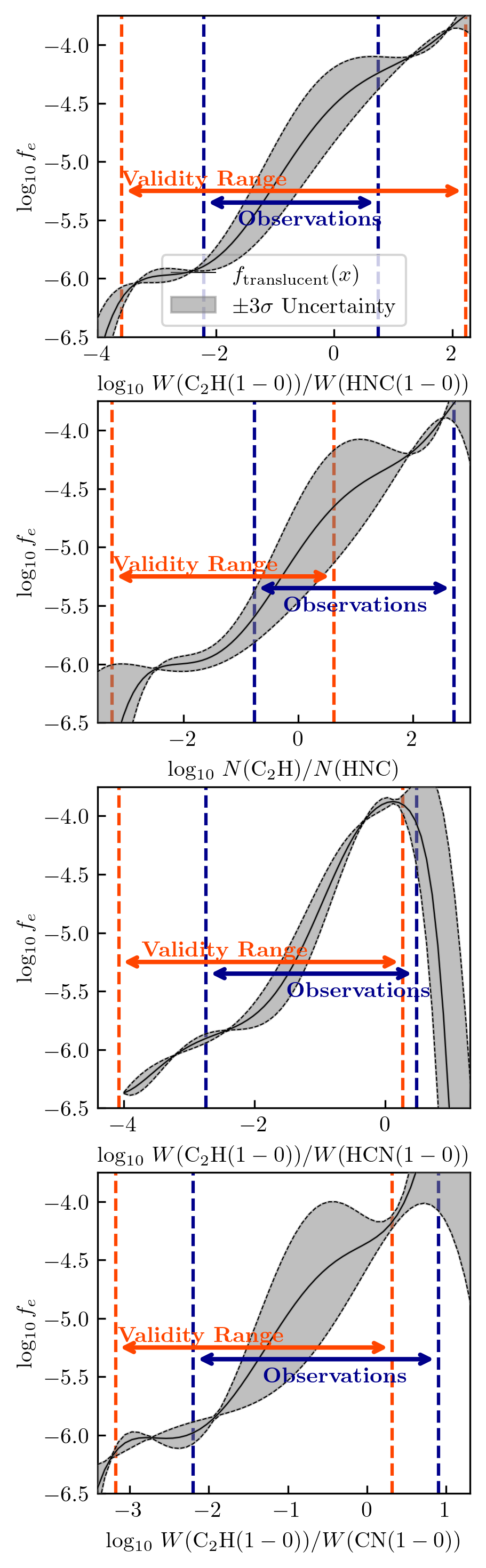}
    \caption{The same as in \ref{fig:dense_int_models}, but in the case of translucent gas. Here, we show  \fe\, as a function of intensity ratios of \CCH/\HNC{} (top panel), column density ratio of \CCH/\HNC\, (second panel), \CCH/\HCN{} (third panel), and \CCH/\CN{} (bottom panel).}
    \label{fig:translucent_int_models}
\end{figure}

Fig.\,\ref{fig:dense_int_models} shows the ionization fraction computed for dense medium by applying Eq.\,\ref{eq:xe} and using coefficients from Tab.\,\ref{tab:coeff}. In all panels, the black line represents analytical function (Eq.\,\ref{eq:xe}), and the grey shaded area shows its $3-\sigma$ variance. Purple horizontal line corresponds to the validity range derived from models in B21. The validity range is defined as the range of values in intensity and column density ratios within which the analytical formulae (Eq.\,\ref{eq:xe}) are valid (the uncertainty falls below 2$\%$). The blue horizontal line represents the range in values derived from our observations. The first row shows \fe\, derived from CN/\NtHP\, (intensity and column density ratios), while the bottom row shows \fe\, computed using \thCOHCOp\, and \CeiOHCOp.

The largest overlap with the validity range is observed for the intensity ratio of \thCO/\HCOP{} (bottom right panel), where we see that the entire validity range was covered in our measurements. Next is the \CN/\NtHP\, intensity ratio (top left panel), where the overlap spans three orders of magnitude. Our observations did not cover values of \CNNtHP\, lower than $10^{-1.8}$. 
In addition, only a few lines of sight have an observed \CNNtHP\, higher than the maximum value provided from the validity range. Interestingly, our measurements cover a lower part of the validity range for the column density ratio of \CN\, and \NtHP\. The log of \thCOHCOp\, and log of \CeiOHCOp\, range from -0.25 to 2.25 and -1 to 1.25, respectively. These ranges fall within the validity range for both line ratios, which is somewhat wider, especially on the left side. 

In Tab.\,\ref{tab:validity_ranges} we show the percentages of pixels for each molecular ratio used in this work that fall within the validity range \citep{Bron+2021}. For each ratio considered in dense gas, we observe that most pixels (more than 90 per cent) fall within this validity range.

\begin{table}[]
\caption{The percentage of pixels of each intensity ratio in the model validity range of B21.}
\label{tab:validity_ranges}
\setlength\extrarowheight{6pt}
\begin{tabular}{l|lll}
\hline 
Model                            & $W_1/W_2$ & $N_\mathrm{tot}$ & $N_\mathrm{val}/N_\mathrm{tot}$ [$\%$] \\ \hline \hline 
\multirow{3}{*}{Dense gas}       & $^{12}$CN/N$_2$H$^+$           & 12129             &  94.5                                    \\
                                 & $^{13}$CO/HCO$^+$              & 20218            & 99.5                                   \\
                                 & C$^{18}$O/HCO$^+$              & 20218            & 99.2                                   \\ \hline 
\multirow{3}{*}{Translucent gas} & C$_2$H/HNC                     & 52674             & 100                                     \\
                                 & C$_2$H/HCN                     & 79839             & 99                                   \\
                                 & C$_2$H/CN                      & 46415             & 91      \\ \hline                         
\end{tabular}
\end{table}

\subsection{Translucent medium}
\label{app:translucent}

For the case of translucent gas, we show \fe\, as a function of line ratios in Fig.\,\ref{fig:dense_int_models}. \CCH/\HNC{} intensity ratio derived from observations shows the largest range compared with other line ratios considered in this work, and it takes  values in the validity range. The observed \CCH/HCN{} takes the lower values of the validity range. In the case of \CCH/\CN{}, this line ratio shows the smallest overlap with the validity range, as well as \CCHHNCcd.

\section{Ionization fraction computed from column density ratios} \label{app:cd_results}

\begin{figure}[t!]
    \resizebox{\hsize}{!}{\includegraphics{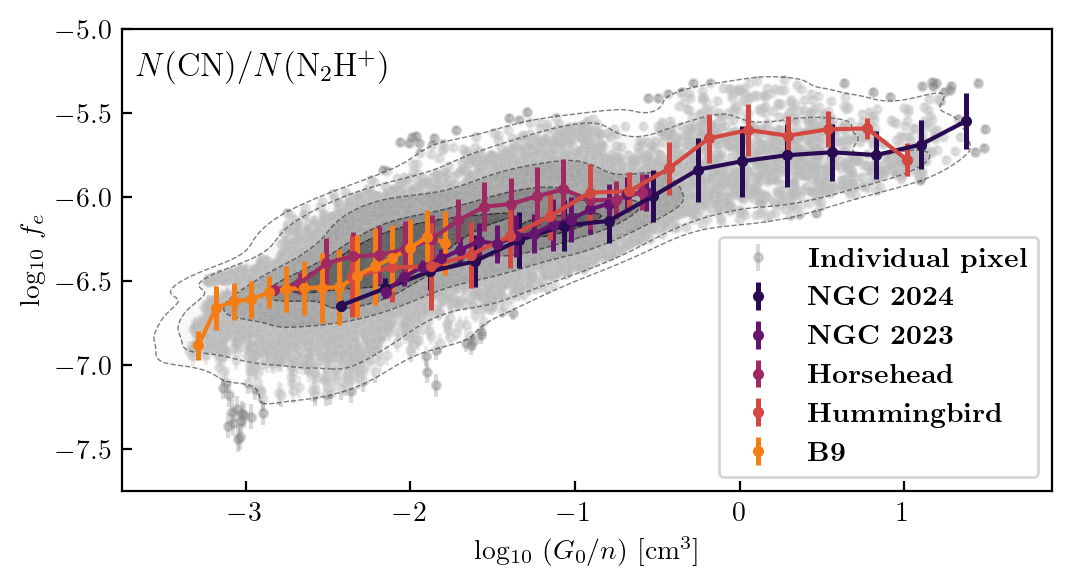}}
    \caption{The same as in Fig.\,\ref{fig:cn_n2hp_G0_G0n}, but using \CNNtHPcd\, to compute the ionization fraction.}
    \label{fig:cn_n2hp_G0_G0n_cd}
\end{figure}

\begin{figure}[hbt!]
    \resizebox{\hsize}{!}{\includegraphics{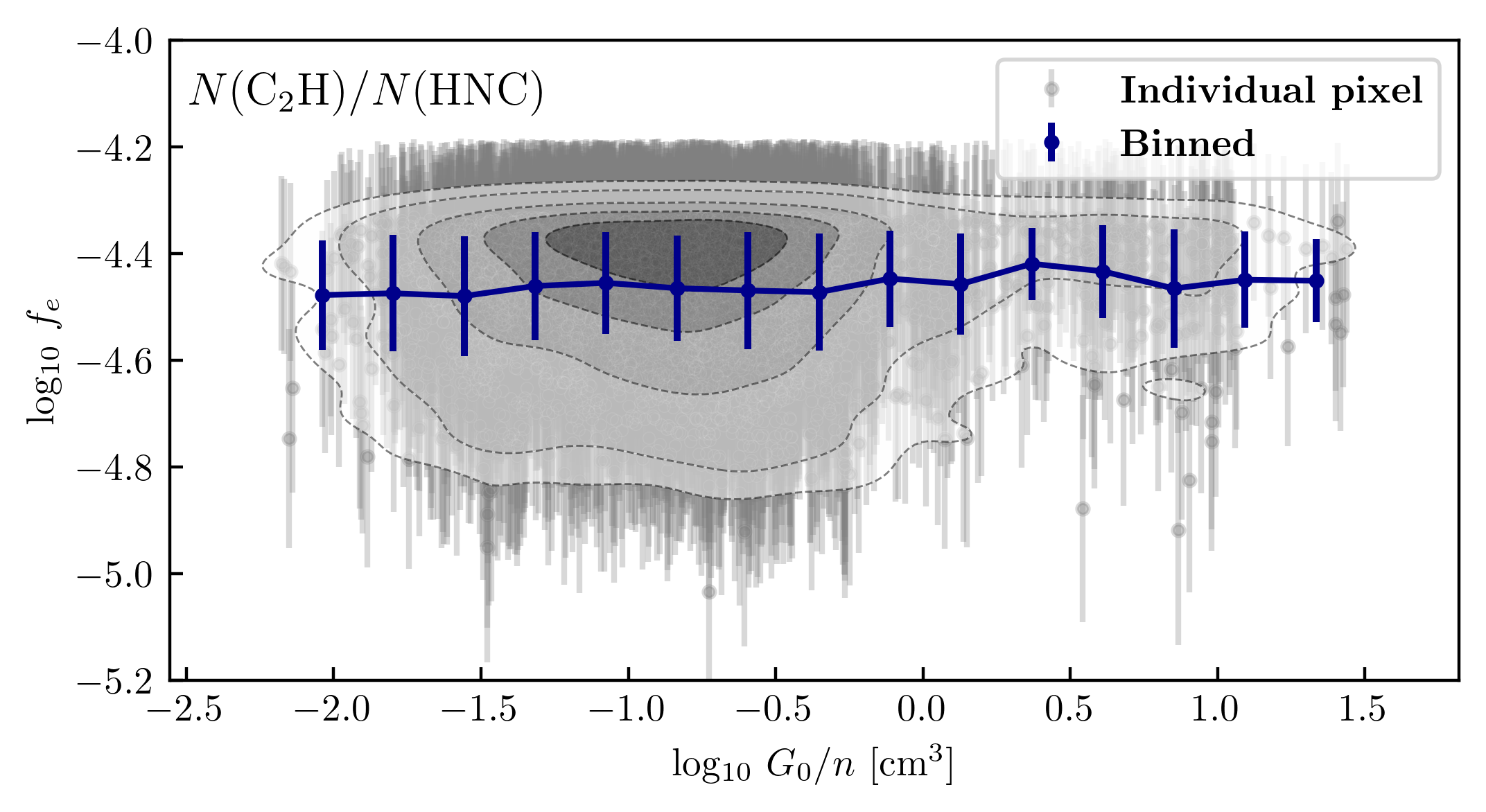}}
    \caption{The same as in Fig.\,\ref{fig:cch_hnc_n}, but we show  \fe\, computed from the column density ratio of \CCHHNC.}
    \label{fig:cch_hnc_cd}
\end{figure}

\subsection{The ionization fraction in dense gas from the column density ratio of \texorpdfstring{CN/N$_2$H$^+$}\,} \label{subsec:dense-cd}

We present results on \fe\, in dense gas computed from column density ratios of CN and \NtHP. We emphasize that we use a different approach to compute column densities than the one used in B21, where column densities were determined from the chemical models. After the column densities are derived, we use Eq.\,\ref{eq:xe} and the corresponding coefficients from Tab.\,\ref{tab:coeff_cd} to compute \fe. All of \CNNtHPcd\, pixels fall within the validity range (see the second panel in Fig.\,\ref{fig:dense_int_models}) defined in B21. 

We show results on the ionization fraction derived from the column density ratio of CN and \NtHP\, as a function of \Gon\, in Fig.\,\ref{fig:cn_n2hp_G0_G0n_cd}. As in Sec.\,\ref{subsec:dense-intensities} and Fig.\,\ref{fig:cn_n2hp_G0_G0n}, we show here binned trends of the ionization fraction of each region. Using a column density ratio as a tracer of the ionization fraction, we find similar values of \fe\, than those derived using the intensity ratio of the same molecular species. These results imply that our assumption of the LTE and optically thin emission to compute column densities is valid in the dense gas in Orion~B. The scatter in Fig.\,\ref{fig:cn_n2hp_G0_G0n_cd} is smaller than in the case of \fe\, computed from intensity ratios. The ionization fraction changes from 10$^{-7.2}$ to 10$^{-5.5}$, and increases toward regions with high \Gon. The lowest ionization fraction is found in low \Gon\, parts of the Hummingbird and across B~9, whereas the highest \fe\, is found across NGC~2024.

\subsection{Ionization fraction from the column density ratio of \texorpdfstring{\CCH(1-0)/\HNC(1-0)}\,} \label{subsec:trans-cd}

The ionization fraction computed from the column density ratio of \CCH\, and \HNC\, as a function of \Gon\, is presented in Fig.\,\ref{fig:cch_hnc_cd}. The observed column density ratio falls more in the right part of the validity range than the \CCHHNC, so the predictions of \fe\, from \CCHHNCcd\, extend to higher \fe\ values. Nevertheless, the median ionization fraction in translucent gas derived from column density ratios is $10^{-4.5}$, consistent with the values derived in Sec.\,\ref{subsec:trans-intensities}.

\section{Ionization fraction in dense gas using \texorpdfstring{\thCOHCOp\,}\, and \texorpdfstring{\CeiOHCOp}\,} \label{app:dense_others}

\begin{figure}[t!]
    \resizebox{\hsize}{!}{\includegraphics{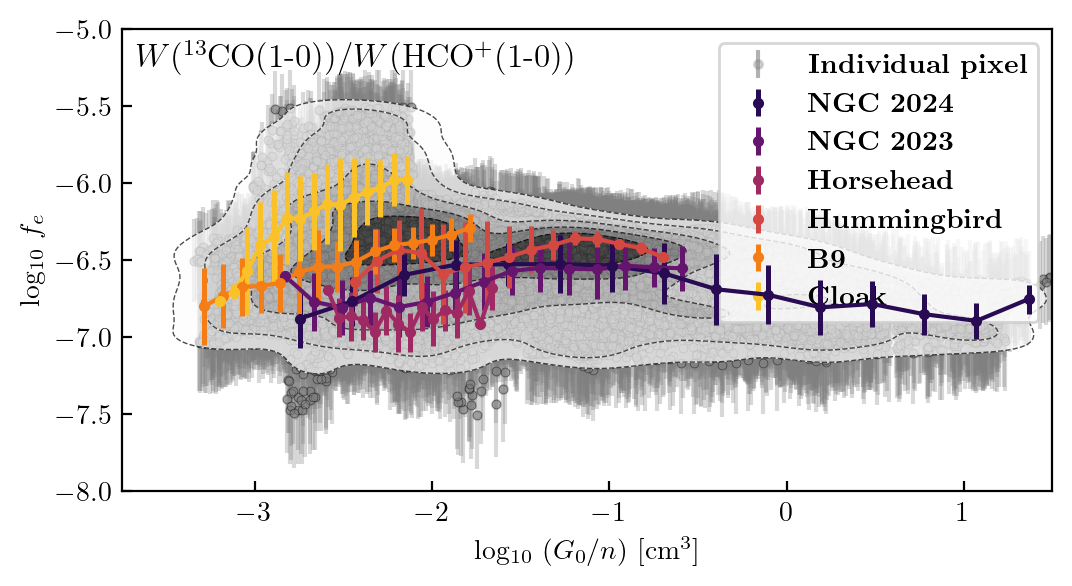}}
    \caption{Same as in Fig.\,\ref{fig:cn_n2hp_G0_G0n}, but for \thCOHCOp.}
    \label{fig:app-13co_hcop}
\end{figure}

\begin{figure}
    \resizebox{\hsize}{!}{\includegraphics{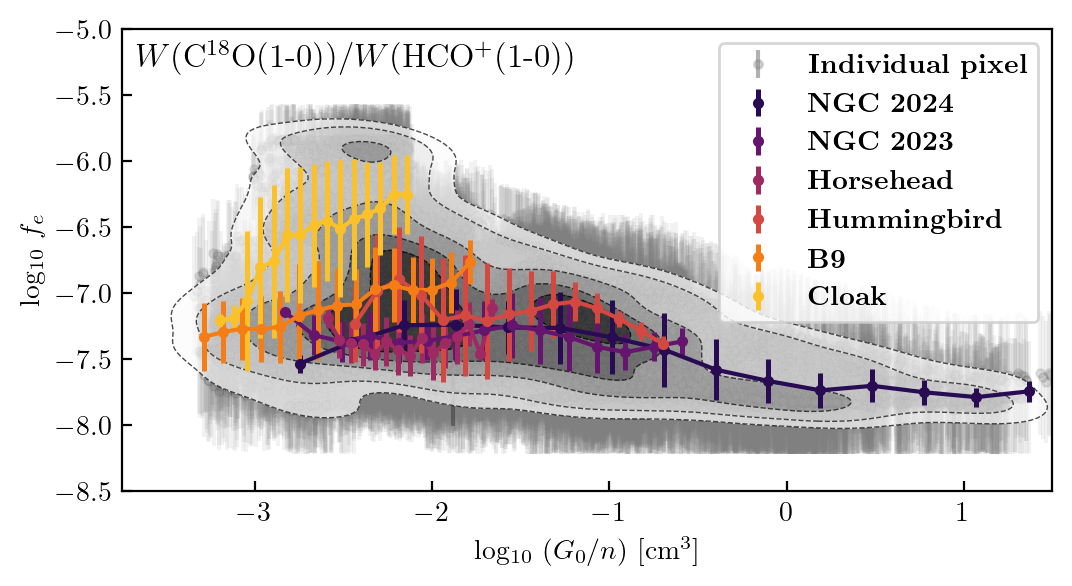}}
    \caption{Same as in Fig.\,\ref{fig:cn_n2hp_G0_G0n}, but for \CeiOHCOp.}
    \label{fig:app-c18o_hcop}
\end{figure}

Moreover, we present measurements of the ionization fraction of Orion~B in dense gas (Sec.\,\ref{sec:dense}) using \thCO/\HCOP\, and \CeiO/\HCOP\, as tracers in Figs.\,\ref{fig:app-13co_hcop} and \ref{fig:app-c18o_hcop} respectively. Similarly, as in Fig.\,\ref{fig:cn_n2hp_G0_G0n}, we show binned trends of each region. In comparison to \fe\, derived from \CNNtHP, we here also detect the target lines in the  Cloak region, whose binned trends are shown in yellow. 

In both these cases, dense regions showing lower radiation field strength, B~9 and the Cloak, have the highest ionization fraction ($\sim 10^{-6.5}$ to $10^{-6}$ from \thCOHCOp, and $\sim 10^{-7}$ to $10^{-6.5}$ from \CeiOHCOp), whereas high-UV illuminated regions show lower ionization fraction, among which the lowest \fe\, is measured toward NGC~2023 and NGC~2024. This supports our conclusion that the line intensity ratios involving \HCOP\, can be biased in regions exposed to intense FUV radiation and provide only a lower limit to the ionization fraction, as already discussed in Sec.\,\ref{subsec:dense_discussion1}.

\section{Ionization fraction in translucent gas using \texorpdfstring{\CCHHCN\,}\, and \texorpdfstring{\CCHCN}\,} \label{app:translucent_others}

\begin{figure}
    \resizebox{\hsize}{!}{\includegraphics{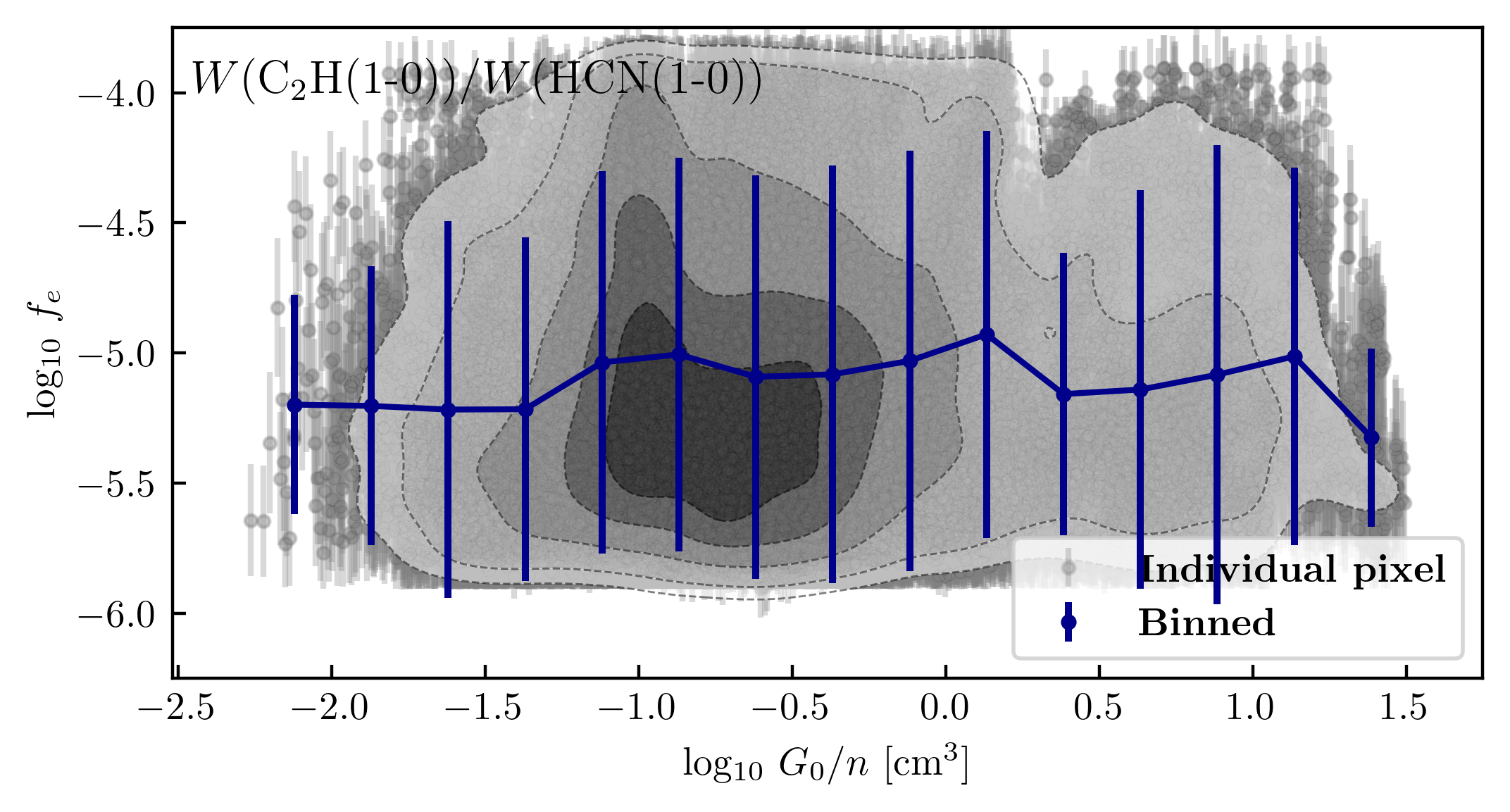}}
    \caption{Same as in Fig.\,\ref{fig:cch_hnc_Gon}, but showing scatter of \fe\, computed from \CCHHCN.}
    \label{fig:app-cch_hcn}
\end{figure}

\begin{figure}
    \resizebox{\hsize}{!}{\includegraphics{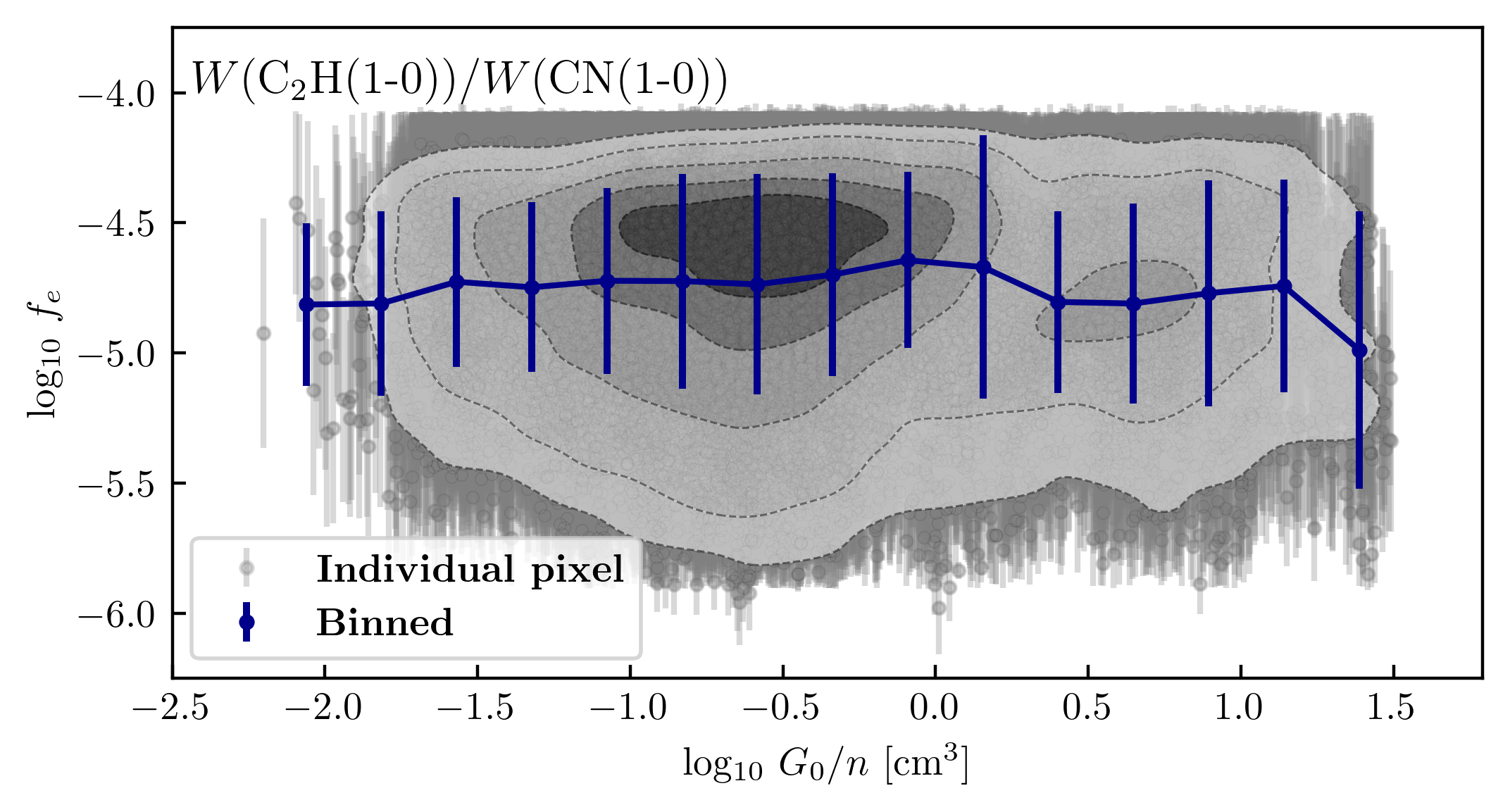}}
    \caption{Same as in Fig.\,\ref{fig:cch_hnc_Gon}, but for \fe\, inferred from \CCHCN.}
    \label{fig:app-cch_cn}
\end{figure}

We show results on the ionization fraction computed from \CCHHCN\, and \CCHCN\, for translucent gas (Sec.\,\ref{sec:translucent}) and corresponding binned trends as a function of \Gon\, in Figs.\,\ref{fig:app-cch_hcn} and \ref{fig:app-cch_cn}. We do not observe significant variation in \fe\, derived from these two line ratios with \Gon. 

\end{document}